\documentclass[acmsmall,10pt,screen,nonacm]{acmart}
\usepackage[yyyymmdd]{datetime}

\usepackage{tabularx}
\usepackage{dcolumn} %
\newcolumntype{d}[1]{D{.}{.}{\#1}}

\usepackage{subcaption}
\usepackage{booktabs}
\usepackage{enumitem}

\DeclareGraphicsExtensions{.pdf,.png,.jpg}

\usepackage{upgreek}

\definecolor{rqline}{HTML}{1F5C8B}
\definecolor{rqfill}{HTML}{EDF3F8}

\AtBeginDocument{%
  }

\setcopyright{none}
\copyrightyear{2026}
\acmYear{2026}
\acmDOI{}
\acmISBN{}
\acmConference{}{}{}
\date{September 11, 2026}

\definecolor{provOpenAI}{HTML}{10A37F}
\definecolor{provAnthropic}{HTML}{D97757}
\definecolor{provGoogle}{HTML}{4285F4}
\definecolor{provMoonshot}{HTML}{5B5BD6}

\newcommand{\providerglyph}[2]{\textcolor{#1}{\small$#2$}}
\expandafter\newcommand\csname providerbadge@openai\endcsname{\providerglyph{provOpenAI}{\bullet}\,OpenAI}
\expandafter\newcommand\csname providerbadge@anthropic\endcsname{\providerglyph{provAnthropic}{\blacksquare}\,Anthropic}
\expandafter\newcommand\csname providerbadge@google\endcsname{\providerglyph{provGoogle}{\blacktriangle}\,Google}
\expandafter\newcommand\csname providerbadge@moonshot\endcsname{\providerglyph{provMoonshot}{\blacklozenge}\,Moonshot~AI}
\newcommand{\providerbadge}[1]{\ifcsname providerbadge@#1\endcsname
  \csname providerbadge@#1\endcsname\else\textbf{[?provider:#1?]}\fi}

\expandafter\newcommand\csname modelmark@gpt\endcsname{\providerglyph{provOpenAI}{\bullet}\,gpt-5.6-luna}
\expandafter\newcommand\csname modelmark@claude\endcsname{\providerglyph{provAnthropic}{\blacksquare}\,claude-opus-4-8}
\expandafter\newcommand\csname modelmark@gemini\endcsname{\providerglyph{provGoogle}{\blacktriangle}\,gemini-3.6-flash}
\expandafter\newcommand\csname modelmark@kimi\endcsname{\providerglyph{provMoonshot}{\blacklozenge}\,kimi-k3}
\newcommand{\modelmark}[1]{\ifcsname modelmark@#1\endcsname
  \csname modelmark@#1\endcsname\else\textbf{[?model:#1?]}\fi}

\newcommand{\providerlogo}[2]{%
  \IfFileExists{figures/logos/#1.pdf}%
    {\raisebox{-0.15\height}{\includegraphics[height=#2]{figures/logos/#1.pdf}}}%
    {\IfFileExists{figures/logos/#1.png}%
      {\raisebox{-0.15\height}{\includegraphics[height=#2]{figures/logos/#1.png}}}%
      {\providerbadge{#1}}}}

\title[An Empirical Study of Human--AI
  Synergy]{Available but Unclaimed: An Empirical Study of Human--AI
  Synergy}

\author{Robin Welsch}
\orcid{0000-0002-7255-7890}
\affiliation{%
  \institution{Aalto University}
  \city{Espoo}
  \country{Finland}}
\email{robin.welsch@aalto.fi}

\author{Michelle Rausch}
\affiliation{%
  \institution{Aalto University}
  \city{Espoo}
  \country{Finland}}

\author{Pascal Knierim}
\affiliation{%
  \institution{University of Innsbruck}
  \city{Innsbruck}
  \country{Austria}}
\email{pascal.knierim@uibk.ac.at}

\author{Thomas Kosch}
\affiliation{%
  \institution{Humboldt-Universit\"at zu Berlin}
  \city{Berlin}
  \country{Germany}}
\email{thomas.kosch@hu-berlin.de}

\author{Jochen Kuhn}
\affiliation{%
  \department{Faculty of Physics}
  \institution{Ludwig-Maximilians-Universit\"at M\"unchen}
  \city{Munich}
  \country{Germany}}
\email{jochen.kuhn@lmu.de}

\author{Albrecht Schmidt}
\affiliation{%
  \institution{Ludwig-Maximilians-Universit\"at M\"unchen}
  \city{Munich}
  \country{Germany}}
\email{albrecht.schmidt@ifi.lmu.de}

\author{Daniela Fernandes}
\affiliation{%
  \institution{Aalto University}
  \city{Espoo}
  \country{Finland}}
\email{daniela.dasilvafernandes@aalto.fi}

\renewcommand{\shortauthors}{Welsch et al.}

\hypersetup{pdfauthor={Robin Welsch, Michelle Rausch, Pascal Knierim, Thomas Kosch, Jochen Kuhn, Albrecht Schmidt, Daniela Fernandes},pdfsubject={Research preprint}}

\begin{document}

\begin{abstract}
People increasingly reason with large language models (LLMs), yet complementary capabilities do not guarantee outperforming both components. In a between-subjects study, participants (N=535) solved a 40-item battery of matrix reasoning, mental rotation, syllogisms, and letter-string analogies, unaided or with GPT-5.6-Luna, Claude Opus 4.8, Gemini 3.6 Flash, or Kimi K3. Each assisted trial required consultation with the model. Each model answered every item alone 100 times under matched elicitation. The assisted–unaided accuracy difference increased with item-level LLM competence. Deference varied across tasks and increased with competence within tasks. Post-advice confidence distinguished correct from incorrect answers less strongly than unaided confidence. In a reference comparison, about half the increase in LLM accuracy carried through to assisted accuracy. How much of that accuracy gain reached participants differed across the models. These findings motivate evaluating LLMs in interaction with humans and designing support for selective deference that preserves independent reasoning.
\end{abstract}

\ccsdesc[500]{Human-centered computing~Human computer interaction (HCI)}

\keywords{human-AI interaction, complementarity, reliance, large language models, reasoning,
decision support}

\maketitle
\hypersetup{pdfauthor={Robin Welsch, Michelle Rausch, Pascal Knierim, Thomas Kosch, Jochen Kuhn, Albrecht Schmidt, Daniela Fernandes},pdfsubject={Research preprint}}

\begin{figure}[tb]
  \centering
  \includegraphics[width=\textwidth]{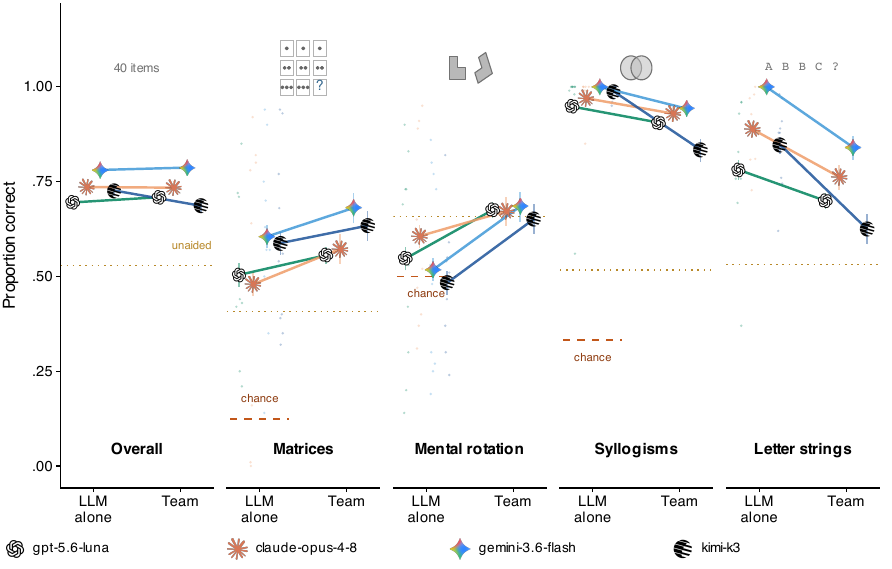}
  \caption{\textbf{Assisted and assistant-alone accuracy differ across tasks.} Each assistant's own accuracy (left, 100-run chat-parity benchmark) and its teams' accuracy (right), overall and per task. The dotted rule marks unaided accuracy and the dashed rule chance (undefined for letter strings). Small points show per-item competences. Bars show 95\% Jeffreys intervals on pooled runs and trials, treating observations as independent.}
  \Description{Five panels, overall and four tasks, with a small schematic of the task at
  the top of each and the task name at its foot. In each panel four lines run from an
  assistant-alone point on the left to a team point on the right, each marked with its
  provider's logo, and a cloud of small points behind the left column shows that assistant's
  per-item competence spread. A dotted line labelled unaided marks the unaided group. A dashed
  line labelled chance marks each task's chance level. The lines are not parallel. The
  assistants that gain most between the two columns are not the ones highest on the left. Thin
  vertical bars behind each mark show 95\% intervals. They are narrow throughout and mostly
  hidden by the marks.}
  \label{fig:teaser}
\end{figure}

\section{Introduction}

Large language models (LLMs) are unevenly reliable within a single domain. A model may solve one reasoning problem on nearly every attempt and a problem of the same form at chance~\cite{Lewis2024,mccoy2024embers}, while its replies need not reveal the difference~\cite{zhou2024reliable}. Whether a person can distinguish correct from incorrect advice and rely on the model accordingly matters for effective human--AI interaction (HAI).

People and LLMs exhibit different strengths and weaknesses when working on a problem. One can theoretically compensate for the other's errors~\cite{hemmer2025complementarity,steyvers_bayesian_2022}. \textit{Complementarity} denotes this potential, created where human and AI errors differ. \textit{Synergy} denotes the realized case in which the team actually outperforms both components~\cite{vaccaro2024combinations}. During an interaction, answers can change as the conversation proceeds. Differing errors therefore create an opportunity, but do not guarantee that the person can recognize and correct them. 
We call the realized share of independent-error reference headroom \emph{synergy capture}.

However, achieving synergy is not built into an LLM's configuration. It depends on the interaction, particularly on what the user understands and how the assistant communicates its reliability~\cite{zhou2024reliable}. That is the familiar concern of reliance, together with its failure mode of overreliance~\cite{doi:10.1518/hfes.46.1.50}. Effective human--AI synergy depends not only on how capable the LLM is, but also on how people use its advice.

LLM assistants are typically reluctant to express uncertainty, and users rely on their output even when it is marked as uncertain~\cite{zhou2024reliable}. Users accordingly underestimate performance variation~\cite{kelly2023capturing} and form biased estimates of model quality~\cite{fernandes2024ai}. Where AI reliance in HAI is easily quantifiable (e.g., in simple decision-making tasks with low to no interactivity), synergy is rare. Meta-analytically, human–AI teams fail on average to exceed the better of their parts in non-conversational interactions~\cite{vaccaro2024combinations}, where per-item correctness is known, reliance can be scored as appropriate~\cite{schemmer2023appropriate}, and complementarity is formally defined~\cite{hemmer2025complementarity}. Interactive studies likewise show that solo LLM accuracy incompletely predicts assisted performance~\cite{chang2025chatbench,riedl2026synergy}.

Four questions organize this paper. Because \emph{LLM competence} is measured per item, we ask \textbf{(RQ1)} where consulting pays and where it costs rather than whether it helps on average. Because we classified each trial's advice, its correctness, and whether it was adopted, we ask \textbf{(RQ2)} whether deference follows that boundary. 
We ask \textbf{(RQ3)} whether confidence supports selective deference and what gains reference policies predict. Because four assistants encountered the same items, we ask \textbf{(RQ4)} whether a more capable LLM yields a stronger team. 

To address these, we report a study in which 535 participants worked the same 40-item battery, spanning matrix reasoning, mental rotation, syllogistic reasoning, and letter-string analogies, either unaided ($N = 187$) or with one of four LLM assistants, namely gpt-5.6-luna ($N = 179$), claude-opus-4-8 ($N = 61$), gemini-3.6-flash ($N = 52$), or kimi-k3 ($N = 56$). Each assistant also answered every item alone over 100 independent runs, using matched stimuli and API settings. This estimates \emph{LLM competence} on the study's own items, with a binomial standard error of approximately .05 or less. %

We find that benefits increase with the LLM's competence on an item, with positive effects resolved in the two upper competence bands. Across the examined contrast, assisted accuracy increased about half as steeply as LLM competence. \emph{Deference} (i.e., submitting the assistant's answer on a given trial) varied across tasks and increased with competence within tasks. Stated confidence distinguished correct from incorrect answers overall, with a modest positive association between discrimination and synergy capture. Retrospective item-level selection estimates a gain of about six points, or two or three questions in forty. Reliability, then, must be measured at the item level, not reduced to a single score for the assistant.

This paper makes three contributions to human–AI interaction.
\begin{itemize}[leftmargin=1.4em,itemsep=2pt,topsep=3pt]
\item \textbf{An account of answer agreement and switching in open-ended chat}, based on conversations classified trial by trial across four assistants. Under mandatory consultation, deference varied substantially by task and also tracked item competence. The own-answer association was uncertain on disjoint trials. These findings motivate examining interaction design, rather than treating inappropriate deference simply as a user deficit.
\item \textbf{Evidence that complementary errors are not enough.} Formal accounts show how metacognitive sensitivity can support beneficial combination under specified conditions~\cite{li2026metacognitive}. Here, confidence distinguished correct from incorrect answers, yet participants still frequently followed incorrect advice. The policy ladder estimates what more selective use of the two components could achieve, distinguishing task- and item-level selection from an oracle that assumes independent errors.
\item \textbf{Assisted accuracy increases less steeply than LLM competence} at the specified contrast. Assistants within a few points of one another alone differed approximately twofold in benchmark-based \emph{pass-through}, the association between item-level LLM competence and assisted accuracy (\autoref{tab:assistants}). The spread was smaller on the realized-advice axis. Together with deference and capture, these slopes extend prior comparisons of solo and assisted performance~\cite{chang2025chatbench}.
\end{itemize}

\section{Related Work}

\subsection{Complementarity in Human--AI Teams}

Complementarity is necessary for human--AI synergy because differing errors create the potential to outperform either component~\cite{bansal_does_2021,hemmer2025complementarity,zoller2025human,steyvers_bayesian_2022}. Empirically, synergy is rare~\cite{vaccaro2024combinations}. A stronger model is a harder baseline to outperform, and overreliance can erode the human contribution until the pair trails the person alone~\cite{klingbeil2024trust}. How much of that complementarity a pair collects depends in part on \emph{metacognitive sensitivity} (i.e., how well confidence separates one's own right answers from one's wrong ones). A Bayes-optimal combination can improve on both components, even with a less accurate assistant, provided accuracy differences and the dependence between human and LLM confidence satisfy the model's conditions~\cite{li2026metacognitive}. This guarantee concerns independently formed judgments combined by a rule. Calibration alone offers no general guarantee either. In binary classification, any deterministic rule combining calibrated probabilities that does not essentially always follow one agent can perform worse than both for some joint distribution of predictions and outcomes~\cite{peng2025nofreelunch}. The benefit of combining predictions also depends on their joint error structure. How complementarity and metacognition support synergy during an open interaction therefore remains an empirical question.

\par Work on model ensembles illustrates why error dependence matters. Models trained on overlapping corpora fail the same cases~\cite{kim2024m}, and systems that share foundation components homogenize decisions~\cite{bommasani2026algorithmic,kleinberg2021algorithmic}. A recent audit of several hundred models reports that two models agree on roughly six of every ten items both answer incorrectly, with error correlation increasing rather than decreasing as models grow larger and more accurate, across architectures and providers~\cite{kim2025correlated}. Routing and ensembling, by contrast, benefit where models genuinely differ in which items they answer well~\cite{jiang2023llmblender,kim2026capable}. These findings concern model--model combinations, but reinforce the need to measure rather than assume complementary errors in human--AI pairs.

Evaluating synergy requires an LLM-alone reference on the study's own items. Major evaluation traditions benchmark assistants in isolation on standard item pools~\cite{legg2007universal,chollet2019measure,srivastava2022beyond,bringsjord2011psychometric,pellert2023ai}. Scores can sit at ceiling~\cite{klein2024performance} or depend on elicitation conditions. ChatBench addresses this mismatch by measuring human-alone, LLM-alone, and assisted accuracy on shared questions, including a free-text benchmark with matched model settings~\cite{chang2025chatbench}. \textbf{We extend this approach to generated reasoning items, relating repeated competence estimates to deference, confidence, and realized synergy.}

\subsection{Reliance and Its Failure Modes}

Complementarity can be lost when accepting advice replaces independent reasoning. 
People accept LLM output even when it is worse than their unaided judgement~\cite{klingbeil2024trust,shekar2024people}. Without corrective feedback, reliance hardens into coarse heuristics about whether to follow the system at all~\cite{lu2021human}, and impressions of its competence persist despite repeated disconfirmation~\cite{colombatto2023illusions}. Interface designs have attempted to restore displaced deliberation with mixed results. Making acceptance a deliberate action reduces overreliance, yet users rate these designs lowest~\cite{forcing2021}. Explanations increase acceptance regardless of advice correctness~\cite{bansal_does_2021,wang2021explanations}. Disclosing weaknesses reduces reliance on incorrect advice only when errors are difficult to detect~\cite{rieger2026error,Qian2024}. The target has accordingly been reframed as \emph{appropriate reliance}, defined as following correct advice while refusing incorrect advice~\cite{schemmer2023appropriate}---with the most effective designs being those that users resist.

Judge--adviser measures score advice-taking as the shift between pre- and post-advice judgements~\cite{bonaccio2006advice,yaniv2004receiving}, with algorithmic as well as human advisers~\cite{logg2019algorithm}. Evaluating \emph{appropriate reliance} and automation bias~\cite{parasuraman1997humans,skitka1999does,wickens2015complacency} additionally requires relating advice-taking to advice correctness on each trial and tracing how answers change during interaction.

Advice-taking measures also miss metacognitive \emph{sensitivity}, or how well confidence distinguishes one's own correct from incorrect answers. Confidence can rise or fall wholesale without improving metacognitive sensitivity, which is why signal-detection accounts were developed to disentangle performance, mean confidence, and metacognitive sensitivity~\cite{rahnev2025comprehensive}, and why increased confidence conveys no information about whether judgment improved. Advice affects both dimensions unequally. An adviser's expressed confidence shifts a judge's own confidence and trust regardless of advice quality~\cite{pescetelli2021role}. 

With an LLM as adviser, confidence can drift toward what the system displays and remain displaced after withdrawal~\cite{li2025confidence}, while offloading can decouple confidence from accuracy~\cite{fernandes2024ai}. Connecting such shifts to performance requires identifying the advice participants actually received and whether they adopted it.

We measure \emph{deference} under required consultation by identifying what the LLM advised, whether the advice was right, and whether the participant submitted it. \textbf{Deference can thus be checked item by item against where following the assistant actually paid.}

Selective deference also requires effort. \emph{Bounded rationality} sets the standard as the best rule available under limited information and time~\cite{simon1955behavioral}. \emph{Rational analysis} asks which policy is adapted to the environment's structure~\cite{anderson1990adaptive}, \emph{resource-rational analysis} prices that policy against the computation it costs~\cite{lieder2020resource}, and \emph{computational rationality} sharpens it to which policy maximises what a person cares about, net of the effort it demands~\cite{oulasvirta2022computational}, so an apparent error may be a correctly priced decision not to spend effort. Offloading is least costly when verification costs exceed its benefits~\cite{risko2016cognitive}, interface designs that facilitate delegation further reduce its cost~\cite{grinschgl2020interface}, and errors introduced by the resource can make that strategy costly~\cite{weis2022know}. These accounts all assume the agent can price the option in front of it, and pricing requires a signal that tracks what that option is worth.
Offloading research shows that people defer based on their own confidence~\cite{boldt2019confidence}, and their decisions approach the optimal policy when confidence is informative~\cite{gilbert2020optimal}.
These accounts motivate comparisons with explicit alternative policies. \textbf{In this study, item-level measurement lets us evaluate reference policies—task- and item-level selection and an independent-error oracle—and compare their conditional gains.}

\subsection{Team Performance and LLM Capability}

These reliance decisions also matter when comparing assistants. One might expect that as LLM competence increases and people defer to the advice, team accuracy would rise accordingly. Which advice people take also depends on the person, interface, and assistant. The complementarity literature characterizes how a team performs relative to its components~\cite{bansal_does_2021,hemmer2025complementarity}, and a large-scale meta-analysis of human--AI performance finds that team performance tracks LLM competence in aggregate~\cite{vaccaro2024combinations}. Note, however, that this trend conceals cases where a stronger AI yields a weaker team, and the mechanism remains unidentified. Work on tuning decision-support systems finds that the match between the assistant's error profile and the user's baseline matters as much as its overall accuracy~\cite{inkpen_advancing_2023}.

Some studies hold the model fixed and vary its stated or observed accuracy~\cite{yin2019understanding}, its error profile at constant accuracy~\cite{inkpen_advancing_2023}, or its explanations~\cite{bansal_does_2021}. Within a single system, higher accuracy does yield larger team gains~\cite{yu2024heterogeneity}. One study pairs clinicians with five models on the same vignettes~\cite{zoller2025human}, evaluating aggregated ranked outputs rather than interactive assistance. ChatBench instead compares two LLMs in human interaction and finds that their solo accuracy gap narrows with assistance~\cite{chang2025chatbench}. Reanalyzing ChatBench, Riedl and Weidmann~\cite{riedl2026synergy} separate individual and collaborative ability using Bayesian item-response models and associate conversational perspective-taking with collaborative performance. Without humans, a model's rank alone and its rank as a helper in a team of LLMs are only modestly related~\cite{wongchamcharoen2026centaurbench}, and combining models stops paying once the single-model baseline is strong~\cite{kim2026capable}.

\par The candidate mechanism is misallocated reliance. Gains from upgrading the LLM depend partly on which advice people follow. Users treat LLM accuracy as steadier across items than it is~\cite{kelly2023capturing}, so a stronger LLM may raise its own bar without improving the team's selectivity. We extend these evaluations by estimating \emph{pass-through}, the association between item-level LLM competence and assisted accuracy, alongside realized synergy and synergy capture. We examine slopes across forty items per assistant rather than relying on the rank order of four assistants. \textbf{Whether a better LLM produces a better team is thus an empirical question, not a given.}

\section{Method}
\label{sec:design}
In the following, we motivate and document our methodological choices. We built and deployed a purpose-made benchmark with a $40$-item cognitive battery spanning four task domains, a web application that administers it under proctoring, and an automated evaluation of state-of-the-art LLMs on the identical items under prompting conditions matched to what participants see.

\subsection{Participants}
\label{sec:method:participants}
A two-sample power calculation for a person-level accuracy difference of $d = .30$ ($\alpha = .05$, two-sided, power $=.80$, \texttt{pwr::pwr.t.test}) gives $n=175.4$, rounded to $176$ per group. The unaided and gpt-5.6-luna groups exceed that target ($N=187$ and $179$). The additional groups ($N=61$ claude-opus-4-8, $56$ kimi-k3, and $52$ gemini-3.6-flash) extend the comparison to other assistants. An item-level proportion based on roughly 50 participants has a maximum binomial standard error of about $.07$, compared with $.04$ for 179 participants. Hierarchical models account for repeated responses and partially pool group estimates.

We recruited through Prolific and redirected participants to our own web application. Each group was a separate Prolific posting, so assignment was by posting rather than randomized within one. Appendix~\ref{app:armcomp} reports per-group demographics and AI experience. The groups are broadly comparable on age, education, and AI-use frequency, with the gemini-3.6-flash group skewing female (67\% vs.\ 45--52\%). Eligible were adults aged 18 or older whose first language was English, who resided in the United Kingdom, and who had not taken part in an earlier batch of the study. The study had to be completed on a desktop computer to avoid excessive scrolling.

Participants gave informed consent through a stepped briefing covering the data collected, storage keyed only to the Prolific identifier with no name or contact details recorded, the browser-based proctoring by AutoProctor, the retention period, and the right to withdraw at any time. Anyone unwilling to be proctored was instructed to return the study on Prolific instead. Assessed against Aalto University's criteria for mandatory ethics review under the Finnish national research-integrity guidelines, the study met none of them (intervention in physical integrity, departure from informed consent, minors without guardian consent, exceptionally strong stimuli, mental harm beyond that of ordinary life, or a security threat) and therefore did not require committee review.

Participants were paid $\pounds 9$ per hour plus performance bonuses~\cite{Bianco2021}.\label{sec:method:motivation} Each group had a separate leaderboard. Its top ten received an additional half of the hourly rate for their session time, and first place received a further $\pounds 100$, split among ties. Participants saw the bonus rules before starting, but no scores or standings were shown beforehand. Completed sessions were paid whether or not their data were retained, except where the session was returned. Returns were requested for incomplete sessions, for participants failing two or more attention checks, and for those the proctoring review indicated had consulted an external AI assistant or been offline for more than 10 minutes. Where a participant contested a return, we paid full or partial compensation rather than pursue it. Proctoring data were used solely for the manual integrity review.

We excluded sessions with fewer than $40$ scored responses or a missing questionnaire section, two or more failed attention checks, or any flag in the AutoProctor manual review, which caught behavior such as consulting an external AI assistant or being offline for more than 10 minutes. Fifty-seven sessions were excluded, thirty because of a technical error and twenty-seven following proctoring review. One of the latter participants had also failed all three attention checks. 

The analyzed sample comprises 535 participants, including 270 men, 260 women, and 5 who reported non-binary or multiple categories or did not disclose, aged 18--76 ($M = 35.2$, $SD = 10.8$), with 71\% holding at least a bachelor's degree. The sample was AI-experienced --- 81\% reported using AI at least weekly and 44\% daily, most often naming OpenAI as their preferred provider (56\%) --- so the reliance we observe is not a novice's first contact with an assistant. The full demographic breakdown by condition is included with the study materials (see the availability note opening Section~\ref{sec:results}).

\subsection{Experimental Design}
\label{sec:method:conditions}

The user study used a between-subjects design with one manipulated factor, assistant availability (\textit{AI} versus \textit{no-AI}). In the \textit{AI} condition, a conversational AI assistant appeared alongside every task. In \textit{no-AI} the battery was identical but the assistant panel was suppressed entirely. Separately from the user study, each assistant was measured alone on the identical items as a reference benchmark (Section~\ref{sec:method:benchmark}). The four assistant groups are described in Section~\ref{sec:method:participants}.

Items were generated from seeded parameters to reduce the risk of using published instruments present in model training data. In \textit{AI}, participants had to consult the assistant at least once per item before submission, but answer controls remained available so they could form an answer first. The study therefore measures mandated consultation, not naturalistic opt-in use. Each group's assistant and low reasoning setting were fixed server-side throughout the session. Google used a 1{,}024-token thinking budget. Benchmark runs used the same settings (Section~\ref{sec:method:benchmark}).

\subsubsection{The item battery}
\label{sec:method:battery}

We selected four reasoning paradigms used in human research and AI evaluation, with tunable difficulty to sample different human and assistant strengths.

\begin{figure}[tb]
\centering
\begin{minipage}[t]{0.42\linewidth}
  \centering
  \includegraphics[width=0.88\linewidth]{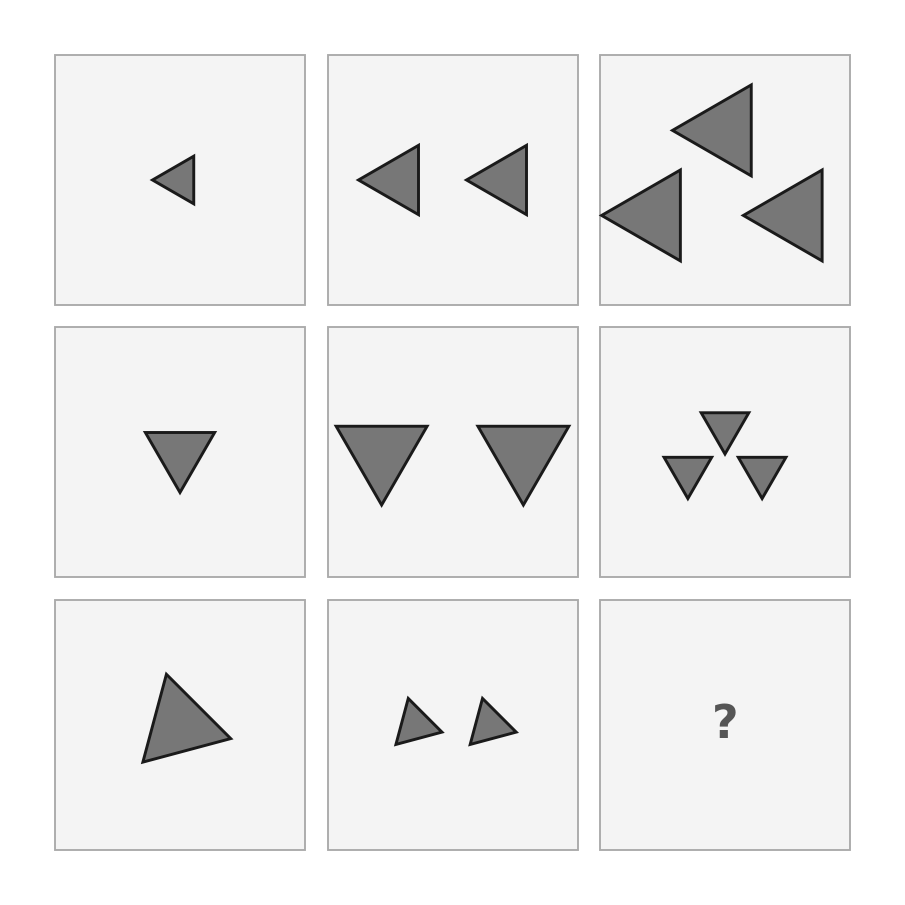}
  \par\smallskip
  \includegraphics[width=\linewidth]{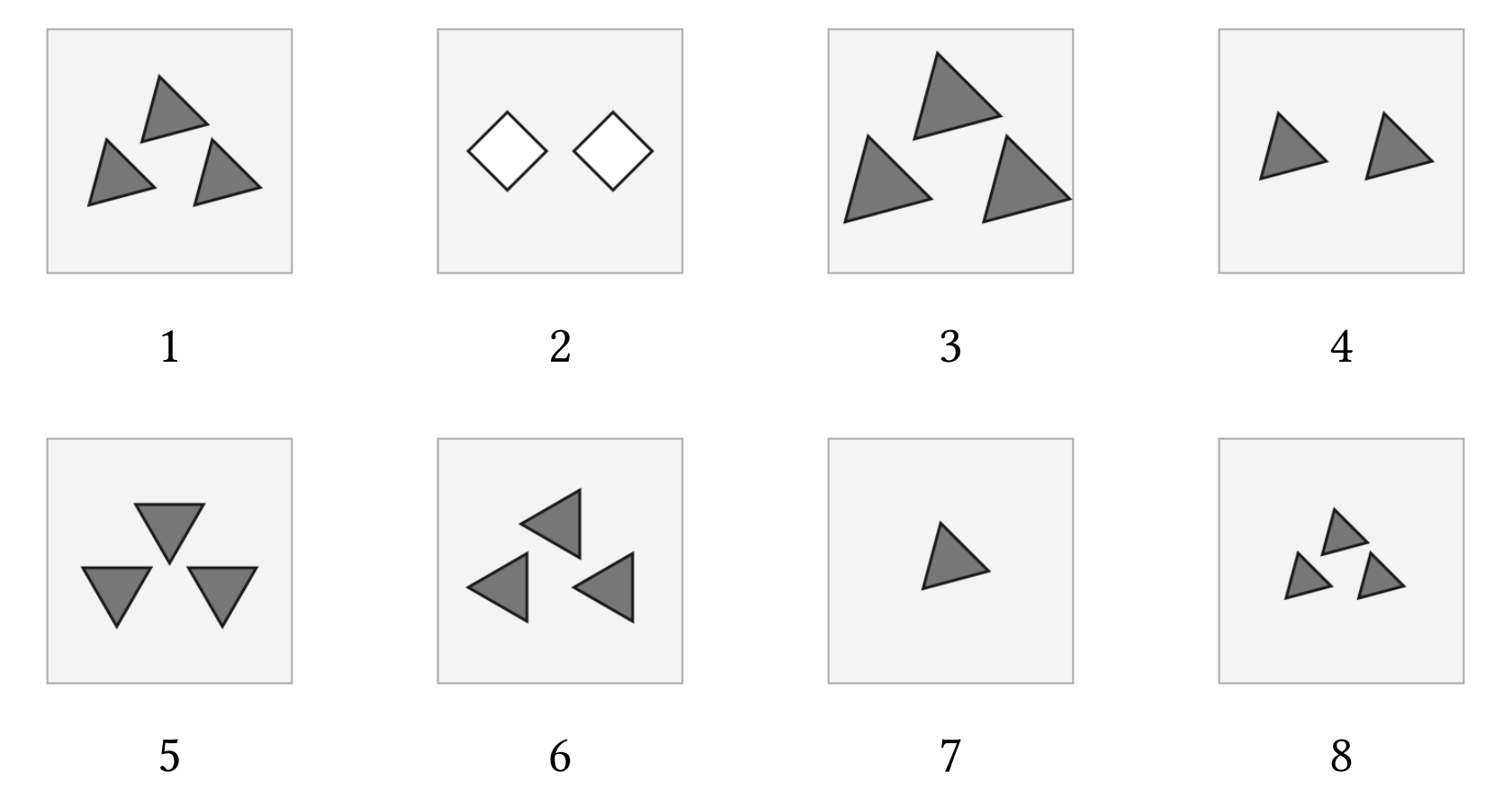}
  \par\smallskip
  {\footnotesize (a) Matrix reasoning. Complete the $3\times3$ pattern using answer options 1--8 below.}
\end{minipage}\hfill
\begin{minipage}[t]{0.53\linewidth}
  \centering
  \includegraphics[width=\linewidth]{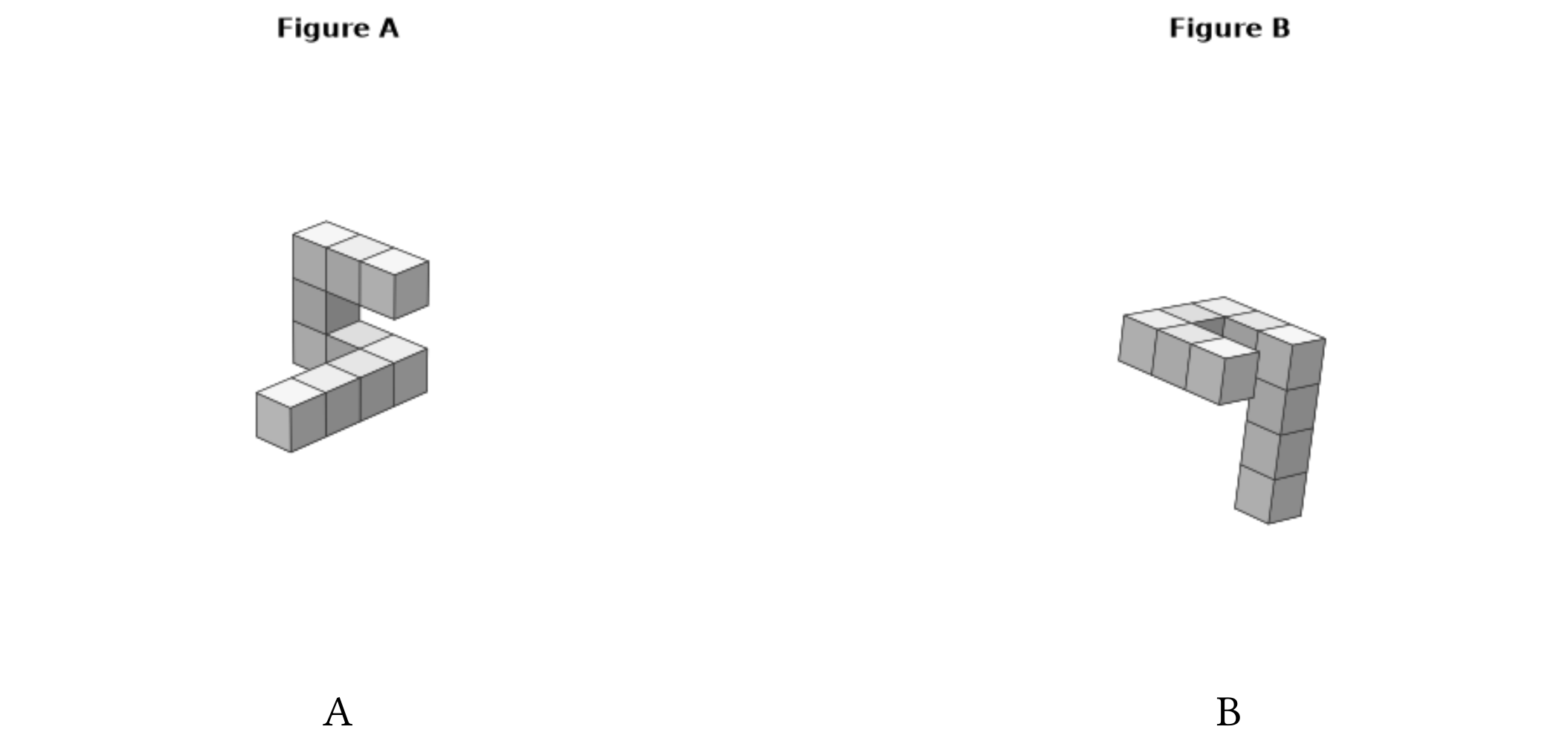}
  \par\smallskip
  {\footnotesize (b) Mental rotation. Is figure~B the same object as figure~A and just rotated, or its
  mirror image?}
  \par\bigskip
  \begin{flushleft}\footnotesize
  (c) Letter-string analogy (free-text answer).\par\smallskip
  \texttt{Alphabet (in order): > * + < ! @ \$ ) \& = : - ( \% \textasciitilde}\par
  \texttt{Study the following patterns:}\par
  \texttt{[@ ) = -] $\rightarrow$ [@ ) = - \%]}\par
  \texttt{[\$ \& : (] $\rightarrow$ [\$ \& : ( \textasciitilde]}\par
  \texttt{Using the same alphabet, complete this:}\par
  \texttt{[> + ! \$] $\rightarrow$ ?}
  \end{flushleft}
\end{minipage}

\par\medskip
\noindent\rule{\linewidth}{0.4pt}\par\smallskip
\begin{minipage}{\linewidth}
\footnotesize (d) Syllogistic reasoning --- the conclusion is judged \emph{Valid} /
\emph{Invalid} / \emph{Cannot be determined}.\par\smallskip
\footnotesize
\emph{Excerpt (record-keeping filler omitted).}\par\smallskip
``At Dunmore Bakery, the manager keeps a daily log of all baked goods and their status. [\ldots]\par\smallskip
Whenever an entry in the bakery's records is classified as items logged by the opening team, it is also classified as items assigned to batch code 3.\par\smallskip
Among the bakery's records, there are entries classified as items assigned to batch code 3 that do not carry the items recorded under the seasonal range code classification. [\ldots]\par\smallskip
\emph{Conclusion:} All items logged by the opening team are items recorded under the seasonal
range code.''
\end{minipage}
\par\smallskip\noindent\rule{\linewidth}{0.4pt}
\caption{One example item per task type; (d) is excerpted for legibility. (a)~matrix item \texttt{mat\_025} (correct answer option~1), (b)~mental-rotation item \texttt{rot\_014} ($80^{\circ}$ depth rotation, correct answer \emph{same object}), (c)~letter-string item \texttt{ls\_014} (correct answer \texttt{> + ! \$ \&}), and (d)~syllogism item \texttt{syl\_013} (correct answer \emph{cannot be determined}). The premises in~(d) are embedded in record-keeping filler so the quantifiers must be inferred from prose.}
\Description{Four example items, one per task type. (a) A matrix-reasoning item shows a three-by-three grid of cells whose triangles vary in number, size, and orientation, with the final cell replaced by a question mark. Eight answer options appear below in two rows, numbered 1--4 and 5--8. (b) A mental-rotation item shows two three-dimensional shapes made of cubes side by side, labelled A and B, with B rotated in depth. The response options are same object or mirrored object. (c) A letter-string item shows two example transformations of short symbol strings followed by a new string whose transformed form must be typed as free text. (d) A syllogism item shows a short workplace narrative containing quantified statements, followed by a conclusion to judge as follows, does not follow, or cannot be determined.}
\label{fig:example-items}
\end{figure}

Seeded offline generators produced a frozen battery with documented item parameters and answer keys (Appendix~\ref{app:generators}). Table~\ref{tab:battery} gives its composition and Figure~\ref{fig:example-items} shows one item per task.

\begin{table}[tb]
\centering
\caption{Composition of the administered 40-item battery. Difficulty labels are
the generator's designed levels, not empirical difficulties.}
\label{tab:battery}
\small
\begin{tabular}{l r l l}
\toprule
Task type & $n$ & Difficulty mix & Response format \\
\midrule
Matrix reasoning        & 10 & 2 medium, 4 hard, 4 expert & Multiple choice, 8 options \\
Mental rotation         & 10 & 2 medium, 4 hard, 4 expert & Multiple choice, 2 options \\
Syllogistic reasoning   & 10 & 1 easy, 4 medium, 5 hard\textsuperscript{a} & Multiple choice, 3 options \\
Letter-string analogies & 10 & 3 hard, 4 expert, 3 combination & Free text entry \\
\bottomrule
\end{tabular}

\vspace{2pt}
\footnotesize\textsuperscript{a}Presented as 6 scenarios, two carrying three questions each and four carrying one.\end{table}

Matrix reasoning follows Raven-like rule designs~\cite{carpenter_what_1990,Matzen2010}, with related items studied psychometrically~\cite{harris_measuring_2020} and used in AI benchmarks~\cite{pmlr-v80-barrett18a,Zhang_RAVEN}. Participants complete a $3\times3$ grid of geometric figures by choosing its missing bottom-right cell from eight images (Appendix~\ref{app:gen:mat}).

Mental rotation follows~\citet{Shepard1971}. Participants judge whether two three-dimensional cube figures are the same object rotated or mirror images. Related rotation tasks assess spatial ability~\cite{branoff_spatial_2009} and AI performance~\cite{ramakrishnan_does_2024} (Appendix~\ref{app:gen:rot}).

Syllogisms, used in human~\cite{khemlani_theories_2012} and LLM reasoning research~\cite{eisape-etal-2024-systematic,ozeki-etal-2024-exploring}, present two-premise narrative arguments judged \emph{Valid}, \emph{Invalid}, or \emph{Cannot be determined}. The ten questions span six scenarios, with difficulty based on mental-model counts~\cite{JohnsonLaird1984} (Appendix~\ref{app:gen:syl}).

Letter-string analogies build on the Copycat paradigm~\cite{hofstadter1995copycat} and its use in human--LLM comparisons~\cite{Lewis2024,Webb2022}. Participants complete a target sequence from two worked examples, using a fictional alphabet whose ordering is stated in the stimulus (Appendix~\ref{app:gen:ls}).

The answers were scored server-side against the stored key, by whitespace- and case-normalized match for letter strings.
On the two letter-string items whose worked examples leave a second rule defensible, the answer following that narrower rule was also accepted as correct (Appendix~\ref{app:gen:ls}).

\subsubsection{Measures}
\label{sec:method:measures}

Task performance was scored as the total number of correct responses out of $40$ and, per task type, out of $10$. Because the four response formats carry different chance levels --- roughly $0\%$ for free-text letter strings, $12.5\%$ for one-of-eight matrices, $33\%$ for three-way syllogisms, and $50\%$ for the two-way rotation judgment --- each task type is interpreted against its own chance level. 

Before and after the battery, participants estimated their own score out of $40$. Assisted participants also estimated their unaided score and the assistant's solo score. After each task, all participants estimated their own, assistant-alone, and counterfactual scores out of $10$, with the counterfactual referring to the assistance condition they had not experienced. Every scored response carried a confidence rating from $0$ (\emph{Unsure}) to $100$ (\emph{Certain}), with one rating per conclusion in multi-question syllogisms. The post-battery global estimates inform the supplementary belief--performance comparison (Appendix~\ref{app:supp}). Additional global percentile and difficulty judgments are documented in the study materials. Backward navigation was disabled to preserve initial estimates.

Every AI-assisted item also includes the full chat transcript with message counts and timings, from which the advice-quality and reliance measures of Section~\ref{sec:method:reliance} are derived. Every letter-string item includes a mandatory free-text field asking participants to describe the pattern they identified and how they applied it (a wrong answer may reflect item ambiguity rather than a reasoning failure). Furthermore, the closing questionnaire asked for an optional strategy description for each task type.

After the battery, we administered four questionnaire sections. The closing questionnaires asked about AI consultation strategy and behavior on disagreement, perceived helpfulness, trust, and frustration (\textit{AI} only), followed by task feedback, demographics, and AI-use frequency (all participants). Three attention checks were embedded. %
Full item wording is included with the study materials (see the availability note opening Section~\ref{sec:results}). The questionnaire structure and attention checks are summarized in Appendix~\ref{app:questionnaire}.

\subsection{Task and Procedure}
\label{sec:method:task}
Participants completed the $40$ items of the battery (Section~\ref{sec:method:battery}) in a randomized order.  Tasks remained contiguous, and syllogism items were shuffled at the scenario level, keeping each scenario's questions together. 

\begin{figure}[tb]
\centering
 \includegraphics[width=\linewidth]{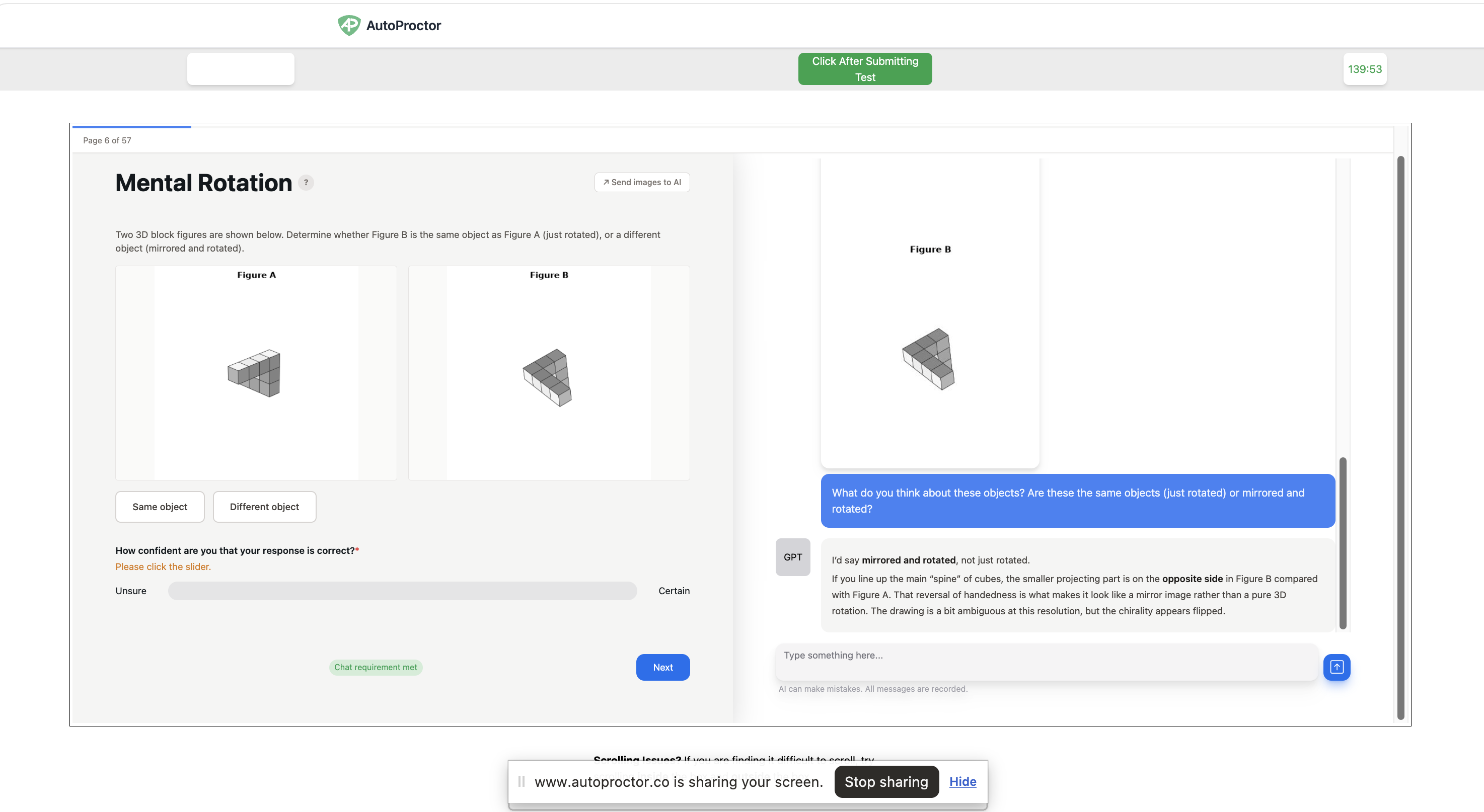}
\caption{The study application used a split interface, with the reasoning task and
the response and confidence controls on the left and the assistant on the right. In
\textit{no-AI} the same pages were shown without the chat panel.} 
\Description{Screenshot of the study web application in the assisted condition. The left side shows the current reasoning item with its answer controls and a progress indicator. The right side shows a chat panel with the AI assistant, a message history, and a text box for sending messages. A header carries the task name and a timer.}
\label{fig:interface}
\end{figure}

The \textit{AI} interface placed a streaming chat panel beside the task (Figure~\ref{fig:interface}). The panel was hidden on questionnaire pages.

\label{sec:method:procedure}After briefing, consent and the pre-task assessment, participants completed four tasks. Each began with instructions and a practice trial, followed by ten scored items and a task-level assessment. Post-task questionnaires and the leaderboard closed the session. Median duration was $86$ minutes (interquartile range $64$--$108$, \textit{AI} $98$, \textit{no-AI} $71$). Screen-by-screen documentation accompanies the study materials (Section~\ref{sec:results}).

\subsection{Apparatus}
\label{sec:method:apparatus}
The study ran in a purpose-built web application. Implementation details accompany the research software (Section~\ref{sec:results}).

AutoProctor monitored screen content, tab switches and navigation away from the study. Sessions flagged in manual review were excluded as described in Section~\ref{sec:method:participants}.

The chat panel provided buttons to copy the stimulus and answer options, excluding confidence and strategy reports, and to upload stimulus images. Both actions were logged.

\subsection{Model Benchmarking}

\label{sec:method:benchmark}
Each assistant was benchmarked on the study's items using matched stimuli and API settings. Unlike participants, assistants received no instruction page or practice trial, and their histories reset between tasks rather than accumulating across the session.

Each assistant completed 100 runs of the 40 items. LLM competence is the proportion of retained replies scored correct per item. With 95--100 replies per cell, the maximum binomial standard error is approximately $.05$ (roughly $\pm.10$ at 95\% confidence). Item order was paired across assistants within each run~\cite{miller_adding_2024}. Shared history makes responses within a run dependent. Runs with API errors were excluded rather than scored.

The models answer in free-form prose, requiring it to be mapped onto the item's answer set before scoring. A second-model extractor, gemini-3.5-flash-lite run with thinking disabled, performs that mapping for every reply. Deterministic normalization and an accepted-alternate table are applied to its output, and for letter strings, a deterministic reading of the reply's own bracketed answer overrules the extractor wherever that answer is unambiguous.

Replies flagged as needing review were excluded. Refusals without that flag were scored as produced. This left 15{,}968 observations. Assistant rankings were unchanged under alternative flag treatments (Appendix~\ref{app:harness}).

\subsection{Coding AI Advice Quality and Reliance}
\label{sec:method:reliance}

We processed each item's complete transcript with a second-model extraction pipeline. Each transcript, stripped of the assistant's reasoning summaries but keeping the turn labels, was passed to gemini-3.5-flash-lite with thinking disabled, instructed to extract, rather than solve for, the first answer the assistant put forward, its last, and the answer the conversation as a whole landed on, each in the item's answer format. For multi-question syllogism scenarios, the conclusion under analysis is named in the prompt. From these we derive whether the final advice matched the key, whether the participant's answer matched that advice, whether first and final differed, and whether the assistant never committed to an answer at all, which counts as incorrect advice but stays distinguishable from a wrong one. Crossing advice quality with whether it was followed and, where it was declined, whether the participant was right, yields the categories used throughout, namely \emph{appropriate reliance}, \emph{costly override}, \emph{overreliance}, \emph{successful} and \emph{unsuccessful override} --- kept apart deliberately, since collapsing the last two would leave ``overrode bad advice'' ambiguous between the participant's own competence and mere disagreement --- and \emph{no advice}.

To check the extractor's followed-versus-declined reading, one author hand-coded 200 transcripts, 25 per group and advice-correctness cell. Nine trials were marked unclear and excluded from the agreement calculation. Agreement on the remaining 191 was 89.5\% ($\kappa = .64$). Task-specific agreement was .97 and .98 on the image tasks, .86 on syllogisms, and .71 on letter strings. The audit and sensitivity analysis concern adoption labels. Advice-correctness extraction requires separate validation. Main intervals treat all extracted labels as fixed.

\subsection{Statistical Approach}

We report posterior medians, 95\% highest-density intervals (HDIs), and directional probabilities $P(b > 0)$. HDIs accommodate skewed posteriors. We call an interval excluding zero \emph{resolved}. One spanning zero leaves the direction uncertain.

\par We estimate descriptive associations on this battery at item, trial, and participant levels. Group contrasts do not isolate effects of access or assistant identity. Item-level models use either 40 item summaries or 160 item--assistant cells. Trial-level models use Bernoulli likelihoods with crossed participant and item effects. Participant-level models use one summary per person. Gaussian models describe confidence, accuracy summaries, and capture. Capture is a ratio modeled on an unbounded scale. Most fixed coefficients have Normal(0, 1.5) priors, with Exponential(1) priors on standard deviations. Intercept and correlation priors vary by model, including brms defaults. The supplementary notebook lists the exact priors for each fit. The benchmark's 100 elicitations per item and assistant are distinct from posterior sampling. The reported Bayesian fits use four chains of 20{,}000 iterations, half warmup. Intervals for aggregated item summaries condition on the estimated component accuracies.

\par Following the Bayesian analysis checklist~\cite{depaoli2017improving} (WAMBS), we audited convergence and prior sensitivity. Across the reported models, the largest $\hat{R}$ was 1.003, the smallest bulk effective sample size 1{,}663, and residual divergences reached 0.49\% of post-warmup draws. Half-width, double-width and flat-prior refits moved probability-scale quantities by at most $.015$, without changing direction or resolution. Models were fit in R with brms (2.23) and cmdstanr. The supplementary notebook gives full diagnostics, sensitivity results and specifications.

\section{Results}
\label{sec:results}

Throughout the Results, \emph{accuracy} and \emph{proportion correct} name the same quantity, the share of items answered correctly, for participants and teams alike. \emph{LLM competence} is the assistant's accuracy on an item, measured over one hundred benchmark runs (Section~\ref{sec:method:benchmark}), and keeps its own name because it serves as the predictor throughout.

\label{sec:method:availability}The study application, item generators, and chat-parity benchmark harness are maintained in \href{https://github.com/RobinWelsch/HAI_Benchmark}{HAI\_Benchmark}. A minimal release of de-identified model inputs, the analysis notebook, the corresponding R script, and reported posterior summaries will be added to the same repository. These inputs support model reproduction rather than reconstruction of raw conversations.
\subsection{Team Accuracy by Item Competence (RQ1)}
\label{sec:rq1}

\begin{figure}[tb]
  \centering
  \includegraphics[width=\linewidth]{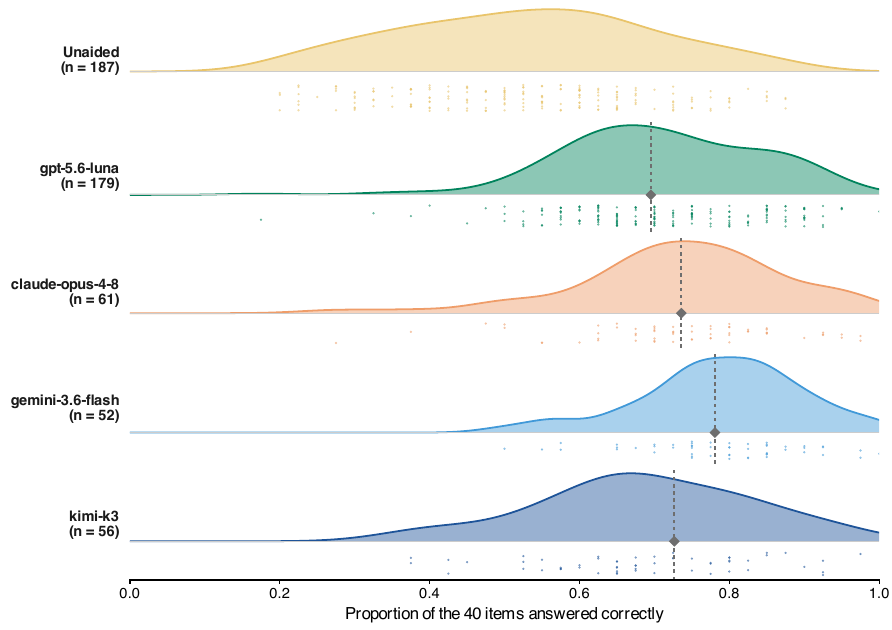}
  \caption{\textbf{Every assisted group outperforms the unaided one, and none exceeds its own assistant by much.} Distribution of individual accuracy over the 40 items, one row per group. \textit{Note.} Each row is a kernel density over participants above jittered points, one per participant. Group sizes are given beside each row. The dashed rule and diamond in each assisted row mark that assistant's own accuracy on the same 40 items without a person in the loop (\autoref{sec:rq4}).}
  \Description{Five stacked rows of individual accuracy, each a density curve above a strip of
  jittered points, one point per participant. The unaided curve is broad and centered near .53.
  The four assisted curves sit further right and are narrower, centered near .69, .71, .73 and
  .79, and their point strips are visibly sparser, the unaided group and gpt-5.6-luna rows carrying
  three times as many participants as the others. A dashed vertical line in each assisted row
  marks that assistant's own accuracy, falling inside the bulk of its own team's distribution in every case except kimi-k3, where it sits above it.}
  \label{fig:density}
\end{figure}
We compare unaided participants, assistants answering alone, and participants working with assistants (teams). Accuracy differences use proportion correct, with $.1$ equivalent to four items. Battery-level comparisons average each component first. Realized synergy $S$ instead compares the team with the better component on each item, a stronger reference. Capture expresses gains relative to an independent-error oracle. Pass-through describes the association between item competence and assisted accuracy.

We now trace through several analyses how \emph{LLM competence} unfolded in performance through interaction.

Every assisted group outperformed the unaided group (\autoref{fig:density}). 
Averaged over items, assisted accuracy exceeds unaided accuracy by $.200$ [$.084$, $.316$], approximately eight of forty items. The comparison with the assistant alone is uncertain, at
$-.006$ [$-.157$, $.140$] (see \autoref{tab:assistants}).

\begin{figure}[tb]
  \centering
  \includegraphics[width=\linewidth]{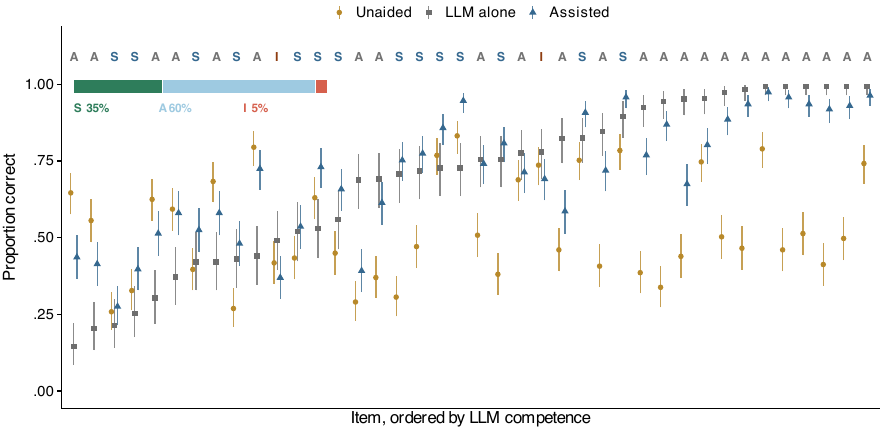}
  \caption{\textbf{Teams track the assistant where it is strong and the unaided participant where the assistant is weak.} Per-item accuracy for the unaided group, gpt-5.6-luna alone, and the group working with it, with the 40 items ordered by LLM competence (the proportion of one hundred benchmark runs answered correctly). Vertical lines are 95\% credible intervals on the item mean. Letters classify each item by point estimates, not resolved differences, with \textbf{S} where it exceeded both of its parts (synergy, 14 items), \textbf{A} where it exceeded one (augmentation, 24), and \textbf{I} where it exceeded neither (interference, 2). The strip at the top left gives the three shares.}
  \Description{Point-and-interval plot of three accuracy series across 40 items ordered by
  LLM competence. The LLM series rises from zero to one across the ordering, the unaided group
  series is roughly flat, and the assisted series follows the unaided group series on the left of the
  plot and the assistant series on the right. A letter above each item classifies the team on that
  item, and a small stacked strip at the top left summarizes the shares, with synergy on 35 percent of
  items, augmentation on 60 percent, and interference on 5 percent.}
  \label{fig:cells}
\end{figure}

We define \emph{realized synergy} $S$ as the team's accuracy on an item minus whichever of its two parts scored higher there, so it is positive only where the pair outperformed \emph{both} of them. Teams beat the unaided participant on 132 of the 160
item-by-assistant cells, and the LLM alone on 61 --- but both at once on only 42
(\autoref{fig:cells}). Averaged over cells, $S$ is negative, $-.062$ [$-.118$, $-.005$], or about two and a half items of the forty.\footnote{%
Excluding the 37 cells where the team answered more than $.90$ of items correctly gives $-.074$
[$-.128$, $-.018$] on the remaining 123, so the negative average is not a ceiling artifact.} $S$ contrasts independent groups. On pre-reply trials, final accuracy minus the higher of the two pooled component accuracies is $-.003$ [$-.019$, $.013$]. This paired estimate pools component accuracies over the selected pre-reply trials, and its direction remains uncertain. Negative average $S$ can therefore coexist with assisted accuracy above the assistant's battery-level average. Per-task results are in \autoref{tab:byblock}.

\emph{Synergy capture} expresses observed gains relative to a complementary reference. The reference is $h+c-hc$, where $h$ and $c$ are unaided-group and assistant accuracy on the item. Capture divides the gain over $h$ by the reference headroom $c(1-h)$.
Unlike accuracy, capture is a ratio. A value of $.500$ means half the reference gain over unaided performance was obtained. Overall capture is $.584$ [$.356$, $.809$], varying across tasks (\autoref{tab:byblock}). The reference assumes independent human and assistant errors. Human--assistant co-failure is measured on trials with an answer committed before the first reply, and therefore inherits that subset's self-selection. On these trials, excess co-failure averages $.036$ [$.010$, $.064$] over the item--assistant cells. Positive dependence lowers this oracle, so capture quantifies reference headroom rather than observed paired potential.

\begin{table*}[tb]
  \caption{\textbf{Realized synergy never resolves positive in any task, capture is uneven across them, and error overlap resolves only on the image-delivered tasks.} Per-task estimates for the three item-level quantities, with tasks ordered by LLM competence. \textit{Note.} $S$ compares assisted accuracy with the better component on each item. Capture is relative to independent-error headroom above unaided accuracy. It is modelled on an unbounded scale, so an interval may extend above 1, and mental rotation leaves almost no headroom, so its ratio is unidentified. Excess co-failure is the amount by which the pair failed together above the product of its two failure rates. Bracketed values are 95\% credible intervals. An interval excluding zero is what we call resolved, and resolved estimates are set in \textbf{bold} throughout the paper's tables. Matrices and mental rotation were delivered as images, letter strings and syllogisms as text. Every task contributes 10 items.}
  \label{tab:byblock}
  \small
  \begin{tabular}{lcccccc}
    \toprule
    & \multicolumn{2}{c}{Realized synergy $S$} & \multicolumn{2}{c}{Synergy capture}
    & \multicolumn{2}{c}{Excess co-failure} \\
    \cmidrule(lr){2-3}\cmidrule(lr){4-5}\cmidrule(lr){6-7}
    Task & Est. & 95\% HDI & Est. & 95\% HDI & Est. & 95\% HDI \\
    \midrule
    Mental rotation & $-.034$          & [$-.100$, .027]    & .071          & [$-.483$, .547] & \textbf{.062}    & [.035, .090]    \\
    Matrices        & .015             & [$-.049$, .078]    & \textbf{.643} & [.383, .888]    & \textbf{.074}    & [.046, .101]    \\
    Letter strings  & $\mathbf{-.153}$ & [$-.218$, $-.090$] & \textbf{.484} & [.293, .689]    & .013             & [$-.014$, .041] \\
    Syllogisms      & $\mathbf{-.074}$ & [$-.139$, $-.009$] & \textbf{.820} & [.641, .997]   & $-.003$          & [$-.031$, .023] \\
    \bottomrule
  \end{tabular}
\end{table*}

\begin{table}[tb]
  \caption{\textbf{Benefits rise with competence, with uncertain effects in the lower competence bands.} The effect function over the 160 item-by-assistant cells, the effect within each competence band, and the slope for each assistant separately. \textit{Note.} The assisted--unaided difference is per-item team accuracy minus unaided accuracy on the same item. Cell-level quantities are on the accuracy scale. The trial-level slope is in log-odds. Resolved estimates are in \textbf{bold}. The zero crossing $-b_{0}/b_{1}$ and the two lowest competence bands are unresolved, so these estimates do not identify a sharp decision threshold.}
  \label{tab:effect}
  \small
  \begin{tabular}{lc}
    \toprule
    Quantity & Estimate [95\% HDI] \\
    \midrule
    \multicolumn{2}{@{}l}{\textit{The effect function}} \\
    Effect at mean competence      & \textbf{.200} [.112, .277] \\
    Slope on competence            & \textbf{.401} [.193, .612] \\
    Effect where the assistant always fails & $\mathbf{-.103}$ [$-.201$, $-.014$] \\
    Zero crossing                  & .238 [$-.153$, .527] \\
    \addlinespace
    \multicolumn{2}{@{}l}{\textit{Effect within competence band}} \\
    LLM right on $\leq$ .2 of runs & $-.020$ [$-.135$, .103] \\
    \quad.2--.5                   & .056 [$-.052$, .158] \\
    \quad.5--.8                   & \textbf{.181} [.076, .283] \\
    \quad $\geq$ .8                & \textbf{.283} [.183, .385] \\
    \addlinespace
    \multicolumn{2}{@{}l}{\textit{Slope, per assistant}} \\
    gpt-5.6-luna (low)             & \textbf{.487} [.378, .597] \\
    gemini-3.6-flash               & \textbf{.447} [.341, .549] \\
    claude-opus-4-8                & \textbf{.382} [.280, .486] \\
    kimi-k3                        & \textbf{.264} [.142, .382] \\
    \bottomrule
  \end{tabular}
\end{table}

Because the four assistants exhibit different item-level competence profiles (\autoref{fig:teaser}), we can estimate how the benefit of access varies with LLM competence by regressing the item-level assisted--unaided difference on the assistant's competence on that item. 

\begin{figure}[tb]
  \centering
  \includegraphics[width=\linewidth]{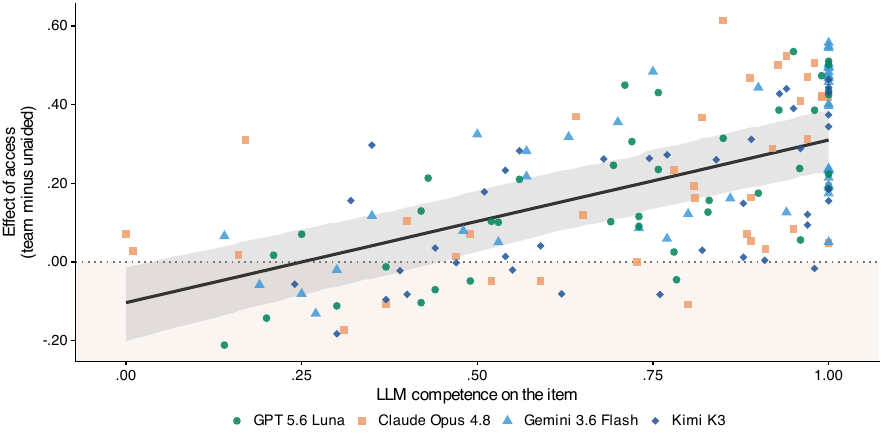}
  \caption{\textbf{Assisted--unaided differences increase with LLM competence.} The 160 item--assistant cells, coloured by assistant, with a pooled fit and 95\% interval. Negative values indicate lower assisted accuracy. The pooled line summarizes the trend; assistant-specific slopes are reported separately.}
  \Description{A scatter plot of the per-item assisted--unaided difference against the assistant's competence
  on that item, with one upward fitted line and its credible band. Points from all four assistants
  are interleaved rather than separated, with no assistant occupying a distinct region.}
  \label{fig:effect}
\end{figure}

The effect function (\autoref{fig:effect}) relates assisted-minus-unaided accuracy to benchmark competence across the 160 item--assistant cells. Its positive slope indicates larger benefits on items the assistant handles better. Positive effects resolve in the two upper bands. The lower bands remain compatible with both benefit and harm (\autoref{tab:effect}). The fitted zero crossing is $.238$ [$-.153$, $.527$], with 91.7\% of posterior mass inside $[0,1]$. The $.50$ boundary defines our reporting bands.\footnote{The task-order adjustment is reported in Appendix~\ref{app:supp}.} Within-task trial slopes are positive in all four tasks, from $2.177$ [$1.525$, $2.830$] in letter strings to $3.011$ [$1.846$, $4.178$] in syllogisms. These slopes describe average trends across items within each task.

\par 
\label{sec:within-task}

To describe differences across observed score levels, we split participants into thirds by battery accuracy within each group, then compared assisted and unaided strata twice, once on the items where their assistant was reliable ($\geq.80$ of its runs correct), and once on those where it was unreliable ($\leq.40$).

\begin{figure}[tb]
  \centering
  \includegraphics[width=\linewidth]{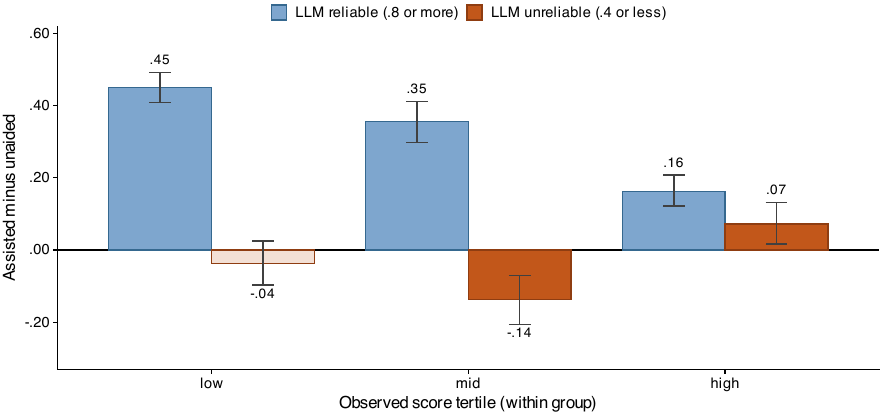}
  \caption{\textbf{Assisted--unaided differences vary across observed score strata and assistant reliability.} Assisted minus unaided accuracy by within-group score tertile, shown separately for items where the assistant is reliable ($\geq.8$) and unreliable ($\leq.4$). \textit{Note.} Tertiles use observed scores within each group, not pretreatment ability. Contrasts come from separate trial-level models for reliable and unreliable items. Bars are 95\% credible intervals, and solid fill marks a contrast whose interval excludes zero. The interval spans benefit and harm only in the lowest score tertile on unreliable items, marked by a hollow bar.}
  \Description{Grouped bar chart of assisted minus unaided accuracy across three score
  tertiles. Reliable-item bars are positive throughout and shrink from low to high scores, from
  about .45 to .16. Unreliable-item bars are near zero and unresolved in the lowest stratum, clearly
  negative in the middle stratum at about $-.13$, and slightly positive in the highest stratum at about
  .07.}
  \label{fig:ability}
\end{figure}

On reliable items, assisted--unaided differences were largest in the lowest score stratum. Gains fall steadily as scores rise, from $+.451$ [$.409$, $.492$] in the lowest third to $+.163$ [$.121$, $.207$] in the highest. Where it was unreliable, the losses do not mirror those gains. It is the middle third that pays, $-.137$ [$-.206$, $-.071$], while the lowest does not resolve and the highest comes out ahead (\autoref{fig:ability}). Refitting one group at a time, the middle-third loss resolves for gpt-5.6-luna, gemini-3.6-flash and kimi-k3, between $-.119$ and $-.160$. No third resolves for claude-opus-4-8. But ``weak items'' means different items for each assistant, drawn from tasks of varying difficulty. Here, the two most dissimilar sets share none.

\subsection{Deference and Answer Switching (RQ2)}
\label{sec:groundtruth}

The extraction pass classified all 13{,}920 assisted trials from 348 participants by extracted advice, its scored correctness, and submitted-answer agreement. On 374 trials the assistant never settled on an answer, leaving 13{,}546 on which there was something to take up. The proportions below are observed trial shares with exact Beta intervals that ignore the clustering of trials within participants and items. Refit as a Bernoulli mixed model with the same crossed random effects as the task models, estimates with participant and item effects set to zero are $.946$ [$.930$, $.960$] for correct advice, $.771$ [$.715$, $.821$] for incorrect, and a discrimination of $.175$ [$.137$, $.217$] --- a different conditional estimand from the observed trial shares.

The extractor identified different first and final answers on .082 of advice-bearing trials, with revisions mostly improving scored accuracy. Of these, 749 moved a wrong first answer to a right one against 236 moving the other way, so three revisions in four improved the answer. Extracted advice accuracy rose from .713 [.706, .721] at the assistant's first answer to .751 [.744, .759] at its last. The revision analysis pools all follow-up types.

Participants submitted answers matching the assistant's advice on .908 [.902, .913]  of the trials where it was correct and on .743 [.730, .756] of the trials where it was wrong, a discrimination of just .165 [.150, .179] (\autoref{tab:tuning}). Incorrect answers were not rare. Participants met a median of nine of them over the battery. A quarter nonetheless never declined a single one. Those who did decline were usually right to, participants who held their own answer against incorrect advice were correct .678 [.658, .701] of the time. 

We use \emph{deference} operationally for submitting an answer matching the assistant's advice, including independent agreement. Recorded answer switches identify replacement of an earlier judgment. Deference varied by task. The four task rates run from .751 on letter strings to .947 on syllogisms, and they do not follow the tasks' advice quality, which itself differs by nearly four tenths (\autoref{tab:tuning}). Letter-string analogies had the second-best advice quality of the four tasks but drew the least deference. The task ordering is sensitive to adoption-coding error.\footnote{An audit-calibrated sensitivity analysis retained frequent wrong-advice adoption ($.787$, 95\% sensitivity interval $[.731,.828]$), but left the letter-string--syllogism contrast uncertain. It assumes coding-error rates transfer within task, advice-correctness and extracted-adoption strata, pooling groups. Advice correctness itself is held fixed.}

Within participants, deference is positively associated with current-task advice accuracy ($b = .093$ [$.035$, $.152$]). The association with previous-task accuracy remains uncertain ($b = .063$ [$-.065$, $.194$]). Holding task fixed, deference also rises with item competence ($b = 1.331$ [$0.983$, $1.695$] in log-odds). Its probability-scale change depends on the task's baseline deference, so it cannot be summarized by a universal four-point change. Both task-level differences and within-task sensitivity contribute to the pattern.

Participants therefore showed some sensitivity to item-level reliability. As a descriptive illustration for the 40-item comparison, deference was .858 [.824, .892] where the assistant outperformed the unaided group and .893 [.833, .950] where it did not. The uncertainty does not support a clear difference, and the unaided-group comparison is not a measure of each participant's own expected accuracy.

\begin{figure}[tb]
  \centering
  \includegraphics[width=\linewidth]{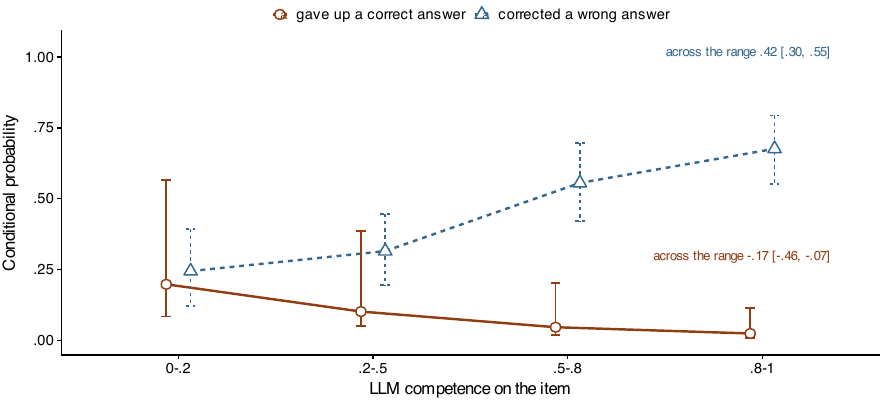}
  \caption{\textbf{Correct-to-incorrect switching is more frequent on low-competence items.} Probability of switching away from a correct answer, and of switching to a correct answer, after the assistant replied, by LLM competence on the item. \textit{Note.} Error bars are 95\% credible intervals from models with crossed random effects for participant and item. Probabilities condition on initial correctness, not on disagreement with advice. The analysis covers the 35\% of assisted trials with an answer recorded before the assistant's first reply. The test each series carries is its change across the competence range, $-.174$ [$-.456$, $-.069$] for abandoning and .425 [.296, .550] for correcting. The two series do not cross at any level of competence.}
  \Description{Two lines across four competence bands. The probability of giving up a correct
  answer falls from about .20 to about .02, while the probability of correcting a wrong answer
  rises from about .24 to about .68.}
  \label{fig:switch}
\end{figure}

Trials with a pre-reply answer show where switching occurred. On the 4{,}873 trials carrying an answer entered before the reply, participants switched from correct to incorrect answers more often on low-competence items --- .198 [.084, .566] of held correct answers in the lowest competence band against .024 [.008, .113] in the highest --- and corrected a wrong one least often there, .244 [.121, .394] rising to .676 [.552, .794] (\autoref{fig:switch}). This pattern is also compatible with differing exposure to incorrect advice.

\subsection{Confidence, Calibration, and the Policy Ladder (RQ3)}
\label{sec:rq3}
 
Confidence concerned the submitted answer and was rated after consultation, including in the pre-reply-answer sensitivity check. An earlier-sample sensitivity analysis includes 1{,}686 rated trials with an answer entered before the reply. The confidence gap is $.052$ [$.030$, $.074$], versus $.105$ [$.095$, $.114$] unaided.

\begin{table}[tb]
  \caption{\textbf{Extracted deference rates differ across tasks.} Advice accuracy, deference and overreliance by task, on the 13{,}546 classified trials that carried committed advice, pooled over all four assistants and ordered by advice accuracy. \textit{Note.} Advice correct is the share of trials on which the assistant's extracted answer was right, and deference the share on which the participant submitted it. Overreliance is deference where the advice was wrong, and discrimination is deference to right minus deference to wrong advice. Task rows are model-based estimates. The pooled row gives observed trial shares. Interval widths are summarized below.}
  \label{tab:tuning}
  \small
  \begin{tabular}{lccccc}
    \toprule
    Task & Trials & Advice correct & Deference & Overreliance & Discrim. \\
    \midrule
    Syllogisms      & 3{,}402 & .969 & .947 & .851 & .100 \\
    Letter strings  & 3{,}384 & .846 & .751 & .482 & .319 \\
    Mental rotation & 3{,}335 & .652 & .902 & .806 & .154 \\
        Matrices        & 3{,}425 & .577 & .867 & .775 & .161 \\
    \addlinespace
    \textit{All tasks} & 13{,}546 & .751 & .859 & .743 & \textbf{.165} \\
    \bottomrule
    \multicolumn{6}{l}{\footnotesize Widest intervals are on advice accuracy in matrices and
    mental rotation ($\pm.12$),}\\
    \multicolumn{6}{l}{\footnotesize which pools four assistants that differ sharply there, then
    $\pm.05$ on overreliance and}\\
    \multicolumn{6}{l}{\footnotesize discrimination in syllogisms, where only 144 trials carried
    wrong advice, and $\pm.04$ on}\\
    \multicolumn{6}{l}{\footnotesize discrimination in letter strings. Every other interval is
    narrower than $\pm.03$.}
  \end{tabular}
\end{table}

\par
\begin{figure}[tb]
  \centering
  \includegraphics[width=\linewidth]{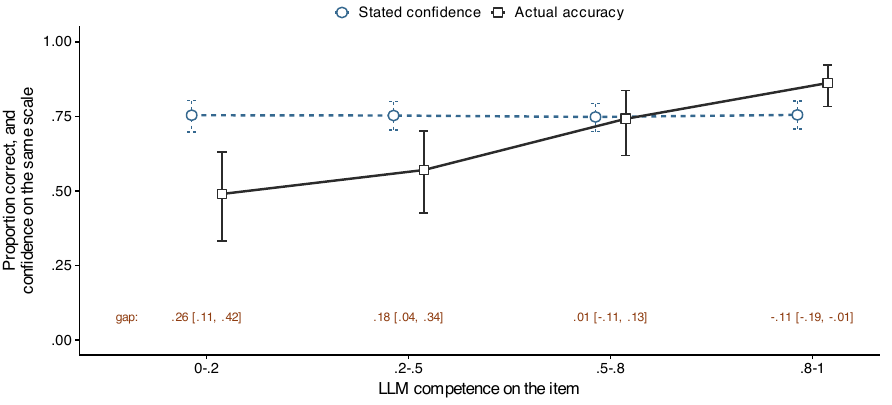}
  
  \caption{\textbf{Stated confidence barely moves while accuracy swings by a third.} Mean stated confidence and realized accuracy by LLM competence on the item, on trials following the intended sequence. \textit{Note.} Post-consultation certainty ratings are divided by 100. Their gaps from accuracy are descriptive scale comparisons, not elicited probability calibration. The figures along the foot give the confidence--accuracy gap in each band, posterior median and 95\% interval. Capped bars on the markers are 95\% credible intervals.}
  \Description{Two lines across four competence bands. Realized accuracy rises steeply from
  about .49 to about .86 while stated confidence stays almost flat near .75, so the vertical
  gap between them closes and then reverses.}
  \label{fig:calibration}
\end{figure}

 Overall, stated confidence was $M = 77.0$, $SD = 25.1$ against an accuracy of .690.
Mean confidence varied little across competence bands while accuracy rose (\autoref{fig:calibration}). The pooled item-level confidence slope is $-.010$ [$-.020$, $.000$], and the within-task estimate is $.000$ [$-.011$, $.012$] (Appendix~\ref{app:supp}). Confidence discriminated correct from incorrect answers overall ($b = .329$ [$.264$, $.396$]). In the weakest band its slope was $.111$ [$-.146$, $.375$], compared with point estimates of $.283$--$.380$ in the other bands. The weak-band interval spans negative and positive confidence--correctness slopes.

\par On the trials that both followed the intended sequence and carried an extracted verdict, confidence was fit on four cells crossing what was submitted (own answer vs.\ LLM's) with whether the advice was right. Among the 11{,}741 adoption trials where participants handed in the assistant's answer, accuracy was 1.000 when the advice was right, and .008 when it was wrong (the residual reflects trials the extractor read as submitted-as-advised that were not, at a rate consistent with the audit in Section~\ref{sec:method:reliance}), yet stated confidence was 78.4 and 78.0. These descriptive means are similar. Submitting the assistant's answer rather than one's own was associated with $11.6$ [$10.3$, $12.8$] points higher confidence, and 48\% of participants were on average \emph{more} confident when the advice was wrong.
In the selected own-answer robustness analysis, capture changes by $-.028$ [$-.140$, $.084$] per SD of discrimination. This association remained imprecisely estimated.

We next examine whether declining wrong advice is associated with metacognitive discrimination or with accuracy on trials where participants retained their own answer. These are observational predictors derived from participants' behaviour during the study.

Metacognitive discrimination was associated with less adoption of wrong advice, with a coefficient of $-.184$ [$-.363$, $-.004$] in log-odds per SD, with an uncertain discrimination interaction. Selected own-answer accuracy was associated with these outcomes in separate models, with coefficients of $-1.018$ [$-1.197$, $-.840$] for taking wrong advice and $1.721$ [$1.559$, $1.889$] for discrimination (\autoref{tab:resist}). These are observational associations. Own-answer accuracy partly overlaps the behaviour being predicted. Its association remained uncertain when source and outcome trials were separated (\autoref{tab:resist}).

Confidence discrimination, measured as the area under the receiver operating characteristic curve (AUC), exceeded .5 in 36 of 40 item-level point estimates, but the slopes relating AUC to synergy and the assisted--unaided difference remained uncertain. AUC counts correct--incorrect confidence pairs, with ties contributing one half. Participant AUC requires both outcomes and at least eight rated trials for deference models, twelve for capture. Participant-level AUC was modestly associated with capture, at $.046$ [$.001$, $.093$] per SD, with $P = .98$ for a positive association. The interval only narrowly excludes zero.

\begin{figure}[tb]
  \centering
  \includegraphics[width=\linewidth]{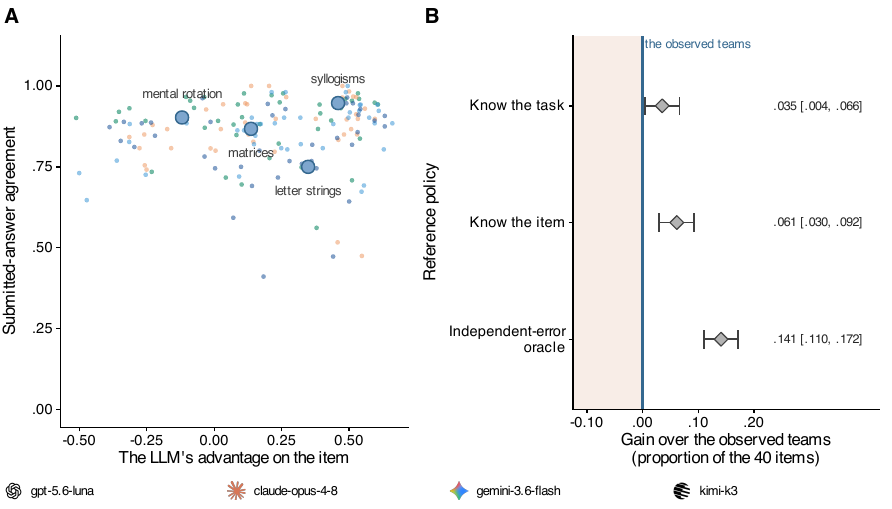}
  
  \caption{\textbf{Answer agreement and reference-policy gains describe different aspects of the interaction.} \textbf{(A)} Submitted-answer agreement against the assistant's item-level advantage, with task means. \textbf{(B)} Three reference policies, each as a gain over observed performance. \textit{Note.} Task-level selection gains $.035$ [$.004$, $.066$], item-level selection $.061$ [$.030$, $.092$], and an independent-error oracle $.141$ [$.110$, $.172$].}
  \Description{Two panels. The left panel plots submitted-answer agreement against the assistant's
  advantage across 160 item--assistant cells, with larger points marking task means.
  The right panel shows three reference rules as gains over the observed
  teams, each with an interval above zero.}
  \label{fig:main}
\end{figure}

The policy ladder compares observed performance with retrospective reference rules (\autoref{fig:main}B). Task- and item-aware references select between estimated unaided-group and benchmark accuracy. Both selection and evaluation use the same data, so their gains may be optimistic.\footnote{Two-fold held-out selection with 2{,}000 participant/run resamples, reselecting each policy, gave task gain $.035$ [$.017$, $.052$] and item gain $.058$ [$.038$, $.074$]. These 95\% percentile intervals condition on the same 40 items.} The trial oracle additionally assumes independent errors. These are conditional reference gains, not demonstrated improvements from deployable policies. The full specification is in Appendix~\ref{app:supp}.

The confidence gap between correct and incorrect submitted answers was lower in every assisted group than in the unaided group (\autoref{tab:confgap}). These are between-group comparisons of post-advice confidence. Mean confidence was negatively associated with competence across items in each group, but the association was carried between tasks and was inconsistent within them (Appendix~\ref{app:supp}).

\subsection{Team Accuracy Across the Four Assistants (RQ4)}
\label{sec:rq4}

\begin{table}[tb]
  \caption{\textbf{Solo and assisted point-estimate rankings differ. Pass-through also varies by group.} Per-assistant solo accuracy, team accuracy, realized synergy, and pass-through, ordered by solo accuracy. \textit{Note.} All values are posterior medians with 95\% credible intervals. Resolved estimates are in \textbf{bold}. Pass-through is the slope of team correctness on the assistant's item competence, in log-odds, partially pooled across assistants. Slopes are group-specific. Realized synergy $S$ compares the team with the better component separately on each item. The \textit{pooled} row is over the 348 assisted participants. The unaided row is shown for reference only. The lower section gives the six pairwise differences in pass-through.}
  \label{tab:assistants}
  \small
  \begin{tabular}{lrcccc}
    \toprule
    Assistant & $n$ & LLM alone & Team & Realized synergy $S$ & Pass-through \\
    \midrule
    gemini-3.6-flash   & 52  & .761 [.674, .845] & .781 [.720, .841] & $\mathbf{-.055}$ [$-.091$, $-.016$] & \textbf{2.81} [2.37, 3.25] \\
    claude-opus-4-8    & 61  & .735 [.653, .816] & .733 [.673, .791] & $\mathbf{-.055}$ [$-.092$, $-.017$] & \textbf{2.21} [1.80, 2.61] \\
    kimi-k3            & 56  & .730 [.648, .812] & .691 [.631, .750] & $\mathbf{-.087}$ [$-.129$, $-.046$] & \textbf{1.39} [.89, 1.88] \\
    gpt-5.6-luna (low) & 179 & .713 [.627, .796] & .711 [.651, .770] & $\mathbf{-.052}$ [$-.088$, $-.013$] & \textbf{2.76} [2.39, 3.14] \\
        \midrule
    unaided            & 187 & ---               & .528 [.474, .583] & ---                                 & --- \\
    \addlinespace
    \textit{pooled}    & 348 & .735 [.633, .834] & .728 [.622, .828] & $\mathbf{-.062}$ [$-.118$, $-.005$] & \textbf{2.46} [2.18, 2.75] \\
    \midrule
    \multicolumn{6}{@{}l}{\textit{Pairwise differences in pass-through} (log-odds)} \\
    \multicolumn{4}{@{}l}{gemini-3.6-flash $-$ kimi-k3}         & \multicolumn{2}{l}{\textbf{1.43} [0.82, 2.03]} \\
    \multicolumn{4}{@{}l}{gpt-5.6-luna $-$ kimi-k3}             & \multicolumn{2}{l}{\textbf{1.37} [.87, 1.88]} \\
    \multicolumn{4}{@{}l}{claude-opus-4-8 $-$ kimi-k3}          & \multicolumn{2}{l}{\textbf{.82} [.29, 1.35]} \\
    \multicolumn{4}{@{}l}{gemini-3.6-flash $-$ claude-opus-4-8} & \multicolumn{2}{l}{\textbf{.60} [.05, 1.13]} \\
    \multicolumn{4}{@{}l}{gpt-5.6-luna $-$ claude-opus-4-8}     & \multicolumn{2}{l}{\textbf{.55} [.13, .98]} \\
    \multicolumn{4}{@{}l}{gemini-3.6-flash $-$ gpt-5.6-luna}    & \multicolumn{2}{l}{.05 [$-.45$, .57]} \\
    \bottomrule
  \end{tabular}
\end{table}

\begin{table}[tb]
  \caption{\textbf{Post-advice confidence separates right from wrong answers less in every assisted group than unaided, and the drop does not scale with what the assistant knows.} The confidence gap between correct and incorrect answers, per group, and each assisted group's drop against unaided. \textit{Note.} The gap is the posterior difference in mean stated confidence (rescaled to $0$--$1$) between correct and incorrect answers, from one model over every rated trial. Drops are differences from the unaided gap on the same scale. Bracketed values are 95\% credible intervals. Every interval excludes zero, so all estimates are set in \textbf{bold}.}
  \label{tab:confgap}
  \small
  \begin{tabular}{lcc}
    \toprule
    Group & Confidence gap & Drop vs.\ unaided \\
    \midrule
    unaided            & $\mathbf{.102}$ [.092, .112] & --- \\
    \addlinespace
    gpt-5.6-luna (low) & $\mathbf{.062}$ [.051, .073] & $\mathbf{-.040}$ [$-.054$, $-.026$] \\
    claude-opus-4-8    & $\mathbf{.057}$ [.037, .075] & $\mathbf{-.046}$ [$-.067$, $-.025$] \\
    gemini-3.6-flash   & $\mathbf{.059}$ [.037, .080] & $\mathbf{-.043}$ [$-.066$, $-.019$] \\
    kimi-k3            & $\mathbf{.053}$ [.035, .072] & $\mathbf{-.049}$ [$-.070$, $-.028$] \\
    \bottomrule
  \end{tabular}
\end{table}

\begin{table}[tb]
  \caption{\textbf{Selected own-answer accuracy and metacognitive AUC are associated with resistance to wrong advice.} Posterior effect of each candidate predictor, per standard deviation of that predictor. \textit{Note.} Overreliance and discrimination are as defined in \autoref{tab:tuning}. Both are trial-level outcomes, so Panel A is in log-odds. The two $n$s give participants in the overreliance and discrimination fits, respectively. AUC coefficients adjust for shrunken own-work ability. Own-answer coefficients are unadjusted. The disjoint estimate uses declined correct-advice trials (median three per person) to predict separate wrong-advice trials. Capture is the realised share of headroom, a participant-level quantity, so Panel B is on the capture scale itself. Bracketed values are 95\% credible intervals, $P$ is the posterior probability that the effect carries the sign of its median, and resolved estimates are in \textbf{bold}.}
  \label{tab:resist}
  \small
  \begin{tabular}{lcc}
    \toprule
    \multicolumn{3}{l}{\textit{Panel A. Who declines wrong advice} (log-odds per $SD$)} \\
    Predictor & Overreliance & Discrimination \\
    \midrule
    Metacognitive discrimination ($AUC$), $n = 320/321$
      & $\mathbf{-.184}$ [$-.363$, $-.004$] & $-.003$ [$-.127$, .117] \\
      & \footnotesize $P = .98$ & \footnotesize $P = .52$ \\
    Own-answer accuracy, $n = 333/334$
      & $\mathbf{-1.018}$ [$-1.197$, $-.840$] & $\mathbf{1.721}$ [1.559, 1.889] \\
      & \footnotesize $P > .999$ & \footnotesize $P > .999$ \\
    Own-answer accuracy, disjoint trials, $n = 283$
      & $-.063$ [$-.283$, $.154$] & --- \\
    \midrule
    \multicolumn{3}{l}{\textit{Panel B. Who captures the headroom} (capture per $SD$, $n = 274$)} \\
    Predictor & \multicolumn{2}{c}{Synergy capture} \\
    \midrule
    Metacognitive discrimination ($AUC$)
      & \multicolumn{2}{c}{$\mathbf{.046}$ [.001, .093], $P = .98$} \\
    Overconfidence, raw
      & \multicolumn{2}{c}{$\mathbf{-.223}$ [$-.261$, $-.185$], $P > .999$} \\
    Overconfidence, own accuracy held fixed
      & \multicolumn{2}{c}{$-.001$ [$-.013$, .010], $P = .59$} \\
    \bottomrule
  \end{tabular}
\end{table}

Four assisted groups are compared against one unaided group. \autoref{tab:assistants} reports solo and assisted accuracy, item-wise realized synergy, and \emph{pass-through}, the association between item competence and assisted accuracy. \autoref{fig:teaser} shows task profiles. Gemini and Opus rank first and second both alone and with participants, while Luna overtakes Kimi with assistance. Luna also has a relatively steep pass-through slope despite its lower solo score. Kimi has the lowest slope on both competence measures. These point-estimate rankings remain descriptive. On the realized-advice axis, the slope spread narrows from roughly twofold to $1.4\times$, and only the gpt-5.6-luna--kimi-k3 contrast resolves (Appendix~\ref{app:supp}). The matched four-assistant benchmark provides a common reference for studying solo and assisted performance.

At the trial level, the pooled competence slope is $2.465$ [$2.182$, $2.750$] in log-odds. For a reference change in benchmark competence from $.25$ to $.75$, the model predicts an accuracy increase of $.263$ [$.222$, $.303$] with participant and item effects set to zero, equivalent to $.526$ [$.443$, $.605$] of the competence change. This roughly half-sized pass-through is specific to that contrast and evaluation convention. The probability-scale fraction varies with baseline competence.

The more competent assistant belongs to the better team on $.776$ [$.701$, $.857$] of item-level pairs that differ.

The task profiles also differ (\autoref{fig:teaser}). Mental-rotation benchmark scores span thirteen points while assisted scores span three, whereas near-ceiling syllogism benchmarks accompany a wider assisted spread. These descriptive comparisons motivate evaluating performance with people alongside performance alone.

\section{Discussion}

We asked when a person and an assistant working together outperform both components. Assisted accuracy fell below the item-wise better-component reference, while the battery-level assisted--LLM difference remained uncertain. An AI assistant is not uniformly reliable. On two items of the same kind, it can be near-certain on one and no better than guessing on the next. We measured that variation by running each assistant repeatedly on the same items. Participants' deference varied substantially by task and also increased with item competence. Consistent with~\citet{vaccaro2024combinations}, we found no clear advantage over the better component. Riedl and Weidmann's modeled AI benefit instead compares assisted with unaided performance~\cite{riedl2026synergy}.

\subsection{Where Consulting Pays (RQ1)}
Complementarity requires that the person and the assistant fail on different questions. Comparing the answer a participant had entered before the reply arrived against the advice that then came yields two separately recorded answers on the same item. Excess co-failure was positive on mental rotation and matrices, while error dependence remained uncertain on letter strings and syllogisms. Thus, complementarity guarantees very little in terms of overall team performance. For letter strings, the assistant was most often reliably correct, although its errors did not always coincide with participants’ errors. Teams, however, scored below the assistant alone. Whether errors differ sets how much a pair could gain. A participant recognizing where the assistant fails may help determine how much they realize.

In line with~\citet{dellacqua2026jagged}, we find assisted--unaided differences vary with item competence. Their jagged frontier runs between whole tasks. Ours runs between items of the same task, which is one level further. Understanding the frontier at that level may help explain a split in the field-experiment literature, where assistance lifts the weakest workers most in some studies and widens the gap in others~\cite{Shakked2023,brynjolfsson2025generative,otis2025uneven}. 

Both can be true, depending partly on the mix of items. An accurate assistant offers more potential gain when unaided accuracy is low. Where reliability varies across items, wrong advice can displace correct answers. The mix of items may therefore contribute to these different outcomes.

\subsection{Deference Varies by Task and Tracks Item Competence (RQ2)}

Li and Steyvers' \cite{li2026metacognitive} model of human–AI synergy, grounded in metacognitive sensitivity, derives conditions under which confidence-informed combination improves on either component. One interpretation of our results is a granularity mismatch. A direct test of the combination rule would require assistant-confidence measurements and implementation of the rule. One possible explanation concerns how each source is weighted by its reliability. LLM competence differed item by item, while deference showed task-level differences. Deference also increased with item competence. The size of that increase depended on the task's baseline deference.

It appears that users represent an assistant as having a single generalizable accuracy rather than multiple~\cite {kelly2023capturing}. Such a belief could favor task-level heuristics, consistent with bounded rationality~\cite{simon1955behavioral,oulasvirta2022computational}. Item-level feedback could help correct that belief. Participants received advice without calibrated item-level reliability estimates. In contemporary LLM interactions, a reply arrives in the same form whether it is right or wrong, and is often overly confident~\cite{zhou2024reliable}, making confidence of expression an unreliable cue to correctness.

This account invites an objection regarding what is rational for the human in interaction with AI. On the ecological-rationality view, a rule is judged not in the abstract but by its fit to the environment it runs in~\cite{payne1993adaptive}, and a simple rule matched to its environment routinely beats an elaborate one~\cite{gigerenzer2009homo}. On the resource-rational view, the cost of the finer rule counts against it~\cite{lieder2020resource}. Following a task-level rule may be reasonable when reliability is hard to judge for each problem. Participants may therefore have adapted to limited information rather than failed to reason. The policy ladder estimates gains from more selective use of advice, assuming the relevant correctness information is available. This estimated potential is what we call “available but unclaimed.” The design challenge is, thus, to help users judge reliability without demanding excessive effort. Whether such support delivers the estimated gains remains to be tested. As assistants become more reliable, routine deference may become more reasonable, while users have less incentive to check the remaining errors.

\subsection{Confidence Alone May Not Supply the Missing Signal (RQ3)}

Our findings motivate testing two approaches, supporting users' assessment of their own answers and providing information about the assistant's reliability. Participants rated confidence in their submitted answers. Post-advice metacognitive discrimination had a modest positive association with capture. The second approach is to display the assistant's reliability on the screen, and here~\citet{rieger2026error} give reason for caution. In their study, explanations were most helpful for items of moderate difficulty, providing the least benefit when errors were obvious and when items were too difficult to evaluate the explanation. Our results likewise suggest examining how item difficulty affects the usefulness of reliability information. Our participants' confidence distinguished their correct from incorrect answers, yet they still frequently adopted incorrect advice. \citet{peng2025nofreelunch} identify a related theoretical limit. Individual calibration need not hold after conditioning on another agent's prediction. Their theorem concerns calibrated binary predictions. It clarifies why the ladder's additional information matters. Task/item selection uses comparative performance, while the oracle assumes independent errors. Whether displayed confidence can support comparable gains is a design question that can be empirically tested.

\subsection{How Solo Competence Relates to Assisted Performance (RQ4)}

Luna's rise above Kimi with assistance, despite its lower solo score, and Kimi's consistently lower pass-through suggest differences in usefulness across the observed groups. The pattern does not fully establish a more ergonomic model. This extends ChatBench's finding that solo-performance differences can shrink in interaction~\cite{chang2025chatbench}. Following~\citet{steyvers_bayesian_2022} and~\citet{hemmer2025complementarity}, we distinguish gains over unaided performance from outperforming both components. On most items, the more competent assistant is also the one whose team does better, on $.776$ [$.701$, $.857$] of the pairs that differ, so the ranking between models largely holds. Benefiting from a better assistant, however, is not the same as outperforming that assistant through selective combination.

An assistant's advantage can be lost either on the way to the user, if a model performs worse in a chat window than on a benchmark, or after it arrives, in what the person does with the reply. We reduced the elicitation mismatch by matching stimuli and API settings. Each assistant received the stimulus message composed by the participants' chat panel, replayed through the same API call with no system prompt or format instruction. However, benchmark histories reset between tasks and omit participant follow-up turns. The gap may, thus,  reflect both elicitation differences and participants' use of advice. 

\subsection{Implications}

\paragraph{For AI development.}
These data compare solo and assisted performance on the same items. Joint evaluation has both conceptual~\cite{haupt2025centaur} and empirical precedents~\cite{chang2025chatbench,riedl2026synergy}, while benchmark construct validity remains a concern~\cite{bean2025construct}. Pass-through adds a measure of how assisted accuracy varies with item competence under a specified protocol. Estimating correctness on new queries requires a proxy for the answer key used here. Repeated sampling offers self-consistency as one candidate whose relation to correctness needs evaluation.

At the worked competence contrast of .25 to .75, predicted assisted accuracy increased by .263. This measures responsiveness, not a fraction of available value lost. Even selecting the better component can yield a half-sized response when unaided accuracy exceeds LLM competence at the lower endpoint. Policy comparisons separately estimate unrealized gains. Interactive evaluation should assess assisted accuracy alongside solo accuracy and users' discrimination between correct and incorrect advice.

\paragraph{For design.}

An answer-first design provides a competing judgment against which to evaluate advice, rather than only a confidence rating. The own-answer association was uncertain when predictor and outcome trials were separated, leaving answer-first design as a hypothesis to test. Making acceptance a deliberate action reduces overreliance, and users rate those designs lowest~\cite{forcing2021}. Eliciting an answer first differs by changing \emph{when} the interface acts rather than adding friction to acceptance—the cost occurs before the reply, during an additional step that may itself require effort. A controlled trial could request a revisable answer or an explicit inability to answer before revealing advice, measuring accuracy, harmful switches, effort, and perceived control. Reliability displays and routing should compare the assistant with the user's available alternative, while preserving the user's final choice. Whether reliance is warranted is not a fixed property of the assistant. It varies with the item at hand.

\paragraph{For augmentation.} These results belong to an old question. \citet{engelbart1962augmenting} claimed that intellect is augmented by a system in which a person, a tool, and a method are adapted to one another, not by the tool alone. Our four assistants differed in the tool and left the method untouched, but solo performance did not translate one-for-one into assisted performance. The findings suggest evaluating methods that help people decide when to follow advice, comparing human judgment against always following the assistant on matched trials, not only against unaided performance.

\subsection{Limitations \& Future Work}
Three limitations bound these results. 

First, results concern four assistants, one interface, and 40 reasoning items under mandatory consultation. Groups were recruited through separate postings, with differing periods and leaderboard sizes. Measured demographics were similar, but unmeasured ability, motivation, and period effects may contribute to group differences. The separate unaided reference estimates group performance rather than each participant's alternative. Optional consultation and longer-term or expert use require further study.

Second, the estimated gains depend partly on uncertainty in LLM competence, whose binomial standard error is at most about $.05$. We assessed this uncertainty by resampling participants and benchmark runs while keeping the 40 items and four groups fixed. Held-out selection checked whether choosing and evaluating policies on the same observations inflated their gains. These checks support our claim of potential gains from more selective advice use on this battery. Achieving those gains in practice remains an open question. The oracle assumes independent errors, and users would need reliability information that is both available and worth the effort of using.

Third, syllogisms provide few incorrect replies, limiting precision for responses to wrong advice. Pre-reply answers are self-selected, and own-answer accuracy uses few, selected trials per participant, including trials contributing to the deference outcome. Randomizing an answer-first requirement would help distinguish the effect of forming an answer from participant selection and shared-data effects. Collecting pre-advice confidence would allow a direct comparison with the post-advice judgments studied here.

\section{Conclusion}
Assisted participants generally outperformed the unaided group but did not clearly exceed their assistants. Gains increased with item competence, while deference reflected both task differences and within-task reliability. Post-advice confidence discriminated correctness, but did not establish a precise account of who captured more of the reference headroom. Retrospective policies estimated additional gains under assumptions about component accuracy and error dependence. These findings motivate testing interactions that support selective deference and independent reasoning, with effectiveness measured against both unaided and assistant-alone performance.

\section*{Use of Generative AI}

We used large language models (Anthropic's Claude Opus 5 and Claude Fable 5.1, OpenAI's Astra with medium reasoning effort, z-ai/glm-5.2, moonshotai/kimi-k2.6, and qwen/qwen3-235b-a22b-2507) to assist with copy-editing
the manuscript and with writing plotting and analysis code. All study design, data collection, statistical analysis, and
interpretation were carried out by the authors, who verified every reported number
against the analysis output and take full responsibility for the content.

\begin{acks}
\small
This work was supported by the European Research Council (ERC) under the European Union's Horizon Europe research and innovation programme, AmplifAI (grant agreement No.~101217557). Views and opinions expressed are however those of the author(s) only and do not necessarily reflect those of the European Union or the European Research Council Executive Agency. Neither the European Union nor the granting authority can be held responsible for them.

Daniela Fernandes is funded by the Finnish Doctoral Program Network in Artificial Intelligence, AI-DOC (decision number VN/3137/2024-OKM-6).

\par\medskip\noindent
\includegraphics[width=0.32\linewidth]{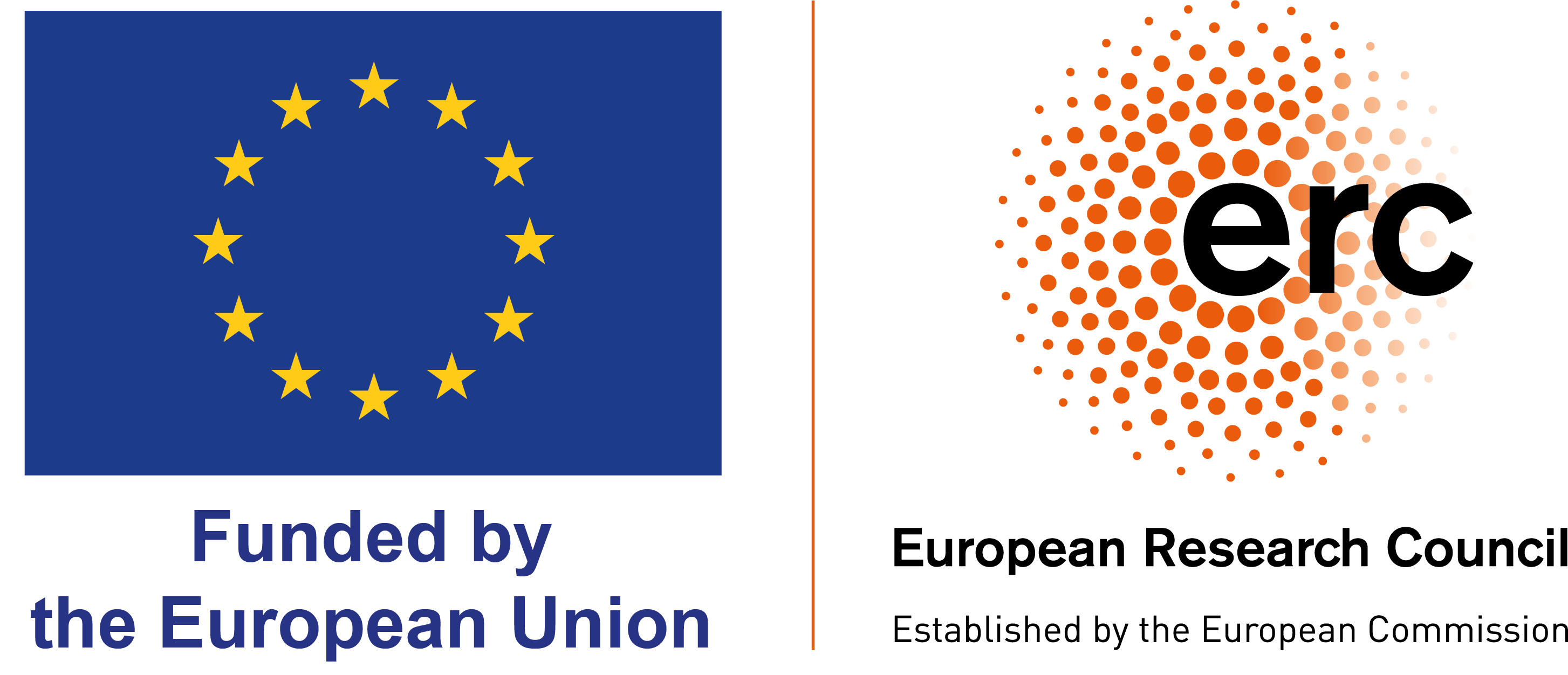}
\end{acks}

\bibliographystyle{ACM-Reference-Format}
\bibliography{bibliography}

\newpage
\appendix
\section*{Appendices}
The appendices document item generation (A), the questionnaire and attention checks (B), the benchmarking harness (C), sample composition (D), and supplementary results (E).
\label{sec:appendix}
\section{Item generation}
\label{app:generators}

\subsection{Matrix reasoning}
\label{app:gen:mat}

Cells vary in shape, size, fill, count and orientation. Each varying attribute is governed by one of four relation types after~\citet{carpenter_what_1990} and \citet{Matzen2010}.

\begin{itemize}
  \item \emph{Constant.} The same value in all three cells of a row, differing
        between rows
  \item \emph{Progressive.} The value cycles across the columns, in the same
        order in every row
  \item \emph{Unique.} The value follows a diagonal cycle, so that the cell in
        row $r$ and column $c$ takes the value at index $(r+c) \bmod 3$
  \item \emph{Distribute-three.} Each of the three values appears exactly once
        per row, in an order drawn at random for each row
\end{itemize}

Difficulty is set by how many of the attributes vary at once and by how many of them use distribute-three, as shown in Table~\ref{tab:matladder}.

\begin{table}[tb]
\centering
\caption{Difficulty levels for matrix items.}
\label{tab:matladder}
\small
\begin{tabular}{l c l l}
\toprule
Level & Attributes varying & Relations drawn from & Additional constraint \\
\midrule
easy   & 2       & classic only & none \\
medium & 3       & classic only & none \\
hard   & 4       & all four     & at least one distribute-three \\
expert & 4\textsuperscript{a} & all four & at least two distribute-three \\
\bottomrule
\end{tabular}

\vspace{2pt}
\footnotesize\textsuperscript{a}At the expert level all five attributes are sampled, but because
orientation is always among them, the shape rule is removed by the orientation--shape exclusion
below and no unused attribute remains to replace it, so every expert item varies the four
non-shape attributes (count, fill, size, orientation), with shape fixed.
\end{table}

Two rendering constraints prevent visually ambiguous options. Orientation rules are restricted to shapes on which orientation is visible (triangles, with the three orientation values separated by at least $25\%$ of the $120^{\circ}$ rotational period). A sampled rule set pairing orientation with a shape rule swaps the shape rule onto an unused attribute, or drops it at the expert level (Table~\ref{tab:matladder}). Distractors are deduplicated on appearance rather than description. A visual key collapses orientation variants that a shape's rotational symmetry renders indistinguishable, and any distractor sharing the correct answer's visual key, or duplicating another distractor's, is removed.

Element size is constant across counts, keeping count and size orthogonal cues, and grid and option images are rendered from shared geometry constants, so a shape is pixel-identical between question and options.

\subsection{Mental rotation}
\label{app:gen:rot}

Figure~A is always shown at its canonical orientation. How Figure~B is produced
depends on the condition. In the \emph{depth} condition the object itself is
rotated in three dimensions. In the \emph{picture-plane} condition the object is
not rotated at all and the camera is rolled by the same angular disparity,
replicating the picture-plane condition of the original paradigm. Both conditions are
crossed with the four difficulty levels in Table~\ref{tab:rotladder}.

\begin{table}[tb]
\centering
\caption{Difficulty levels for mental rotation. The rotation axis applies to the depth condition only, and it switches with the angle, so no axis effect is separable from an effect of angular disparity. With the camera azimuth fixed at $45^{\circ}$ the switch is not itself a difficulty manipulation.}
\label{tab:rotladder}
\small
\begin{tabular}{l c l}
\toprule
Level & Angular disparity & Rotation axis (depth condition) \\
\midrule
easy   & $40^{\circ}$  & world $Y$ \\
medium & $80^{\circ}$  & world $Y$ \\
hard   & $120^{\circ}$ & world $X$ \\
expert & $160^{\circ}$ & world $X$ \\
\bottomrule
\end{tabular}
\end{table}

Every figure is ten cubes in four segments, giving the three right-angled elbows of the original form. Each new cube must be adjacent to exactly one prior cube, which makes closed rings impossible, and base figures are screened for chirality against all 24 proper lattice rotations, so a ``different'' pair is never an achiral figure some rotation could match. Three implementation choices protect the construct. Rotation is applied to polygon face vertices as exact floats rather than integer cube centers, so rotated figures carry no lattice-rounding artifacts. The camera's viewing direction never changes (elevation $25^{\circ}$, azimuth $45^{\circ}$), stationary in the depth condition as in the original method and adding only a roll about the fixed viewing axis in the picture-plane condition. Face shading is reassigned after rotation to the nearest of the six canonical shades by world-space normal, so shading follows orientation rather than identity and cannot signal handedness.

\subsection{Syllogistic reasoning}
\label{app:gen:syl}

Each item is a two-premise categorical argument with a stated conclusion, judged as \emph{Valid} (necessarily true given the premises), \emph{Invalid} (necessarily false), or \emph{Cannot be determined} (neither entailed nor refuted, or the premises are inconsistent). The generator dresses each syllogism as a record-keeping narrative in one of six mundane domains (a bakery, a school, a sports club, a library, an animal shelter, a garden center), rewriting the premises so that the canonical quantifier words (\emph{all}, \emph{no}, \emph{some}) are replaced by prose paraphrases from which the quantifier must be recovered, and weaving in neutral distractor sentences.

Difficulty is computed rather than hand-estimated. Tiers follow the Johnson--Laird mental-model count~\cite{JohnsonLaird1984} --- the number of distinct subject--predicate relationships the premises permit --- computed by an independent solver, replacing hand-typed per-form accuracy estimates that could not be re-derived. The resulting mapping is given in Table~\ref{tab:sylladder}.
\begin{table}[tb]
\centering
\caption{Difficulty levels for syllogism items, and the mean of the hand-typed per-form human-accuracy estimates the computed tiers replaced (shown for validation only).}
\label{tab:sylladder}
\small
\begin{tabular}{l c c}
\toprule
Tier & Mental models & Mean estimated human accuracy \\
\midrule
easy   & 1 & $0.69$ \\
medium & 2 & $0.51$ \\
hard   & 3 & $0.35$ \\
\bottomrule
\end{tabular}
\end{table}
Mean estimated unaided participant accuracy falls monotonically across the three counts, validating the tiers against the estimates they replaced. A fourth tier would need a subjective cut-point, reintroducing the free parameter the computed measure removed.

\paragraph{Multi-question scenarios with disjoint category terms}
Reading a scenario carries a fixed comprehension cost paid once per page, but reusing one premise pair with varied conclusions would make the answers mutually derivable (``All A are C'' exactly negates ``Some A are not C''). A three-question scenario therefore uses nine category phrases in three term-disjoint triples woven into one narrative. No category term is shared between triples, so the premises of one question are logically silent about another's, and disjointness is asserted at generation time. We accepted the price --- a longer page and a ``sorting'' demand --- as a realistic reasoning load.

\paragraph{Balancing response categories during item selection}
All $18$ hard (three-model) single-question items have the answer ``Cannot be determined'' --- a conclusion is valid only if it holds in every model, so the more models a form admits, the less likely any conclusion is entailed --- and difficulty-first selection empirically produced a battery that was $9/10$ ``Cannot be determined'', solvable by a constant response. Difficulty is therefore only a tiebreaker among equally label-useful candidates. ``Invalid'' is scarce because only two of the $48$ forms yield a provably false conclusion, and a hard assertion after selection guarantees at least one valid and one invalid item.

\subsection{Letter-string analogies}
\label{app:gen:ls}
Every item presents the alphabet ordering, two worked examples of one transformation, and one target sequence to complete.
The two alphabets are a permutation of the $26$-letter Latin alphabet in which $20$ of the $26$ positions are deranged (six letters keep their alphabetic position) and a $15$-symbol non-alphabetic set (\textit{> * + < ! @ \$ ) \& = : - ( \% \textasciitilde}). Items are built from four single transformations --- add-letter, fix-alphabet, sort, and a simultaneous successor-and-predecessor operation on opposite ends --- and three ordered pairs of transformations, namely remove-redundant + add-letter, remove-redundant + sort, and sort + add-letter.

Difficulty is built from the generalization axes of~\citet{Lewis2024}, including a step size greater than one, a longer sequence, and grouping, meaning each symbol is displayed twice so the sequence must be parsed before the rule can apply. Each item records how many axes are active at once, counted relative to a baseline of step one, length five and no grouping --- the \emph{medium} level in Table~\ref{tab:lsladder}, which was not administered. 
The alphabet is not one of the counted axes. It changes between levels alongside them --- hard moves to the symbol set as the step rises to two, and expert returns to the $26$-letter alphabet as length and grouping are added --- so no effect of the alphabet can be separated from the axes it moves with.
Table~\ref{tab:lsladder} gives the parameters of each level.

Responses are typed as free text and scored server-side by whitespace- and case-normalized exact match against the key. Runs of whitespace collapse to single spaces and case is ignored, so spacing and capitalization cannot make a correct sequence wrong, and the two documented alternate answers (see below) are also accepted.

\begin{table}[tb]
\centering
\caption{Parameters of the letter-string levels. The \emph{medium} level is
shown as the baseline against which generalization axes are counted, but was not
administered.}
\label{tab:lsladder}
\small
\begin{tabular}{l l c c c c}
\toprule
Level & Alphabet & Step & Base length\textsuperscript{a} & Grouping & Axes active \\
\midrule
medium (baseline) & $26$-letter & 1 & 5 & no  & 0 \\
hard              & $15$-symbol & 2 & 5 & no  & 1 \\
expert            & $26$-letter\textsuperscript{b} & 2 & 9 & yes & 3 \\
combination       & mixed\textsuperscript{c} & 2 & 5 & yes & 3\textsuperscript{c} \\
\bottomrule
\end{tabular}

\vspace{2pt}
\footnotesize\textsuperscript{a}An \emph{element} is one position in the sequence, that is, one symbol of the alphabet. Base length is the length of the sequence the generator draws. The strings actually shown are derived from it, so how many elements they contain depends on the transformation. An add-letter item shows one element fewer before the arrow than after, a remove-redundant item one more. Shown strings run from $4$ to $6$ elements at the medium, hard and combination levels and from $7$ to $10$ at expert. Where grouping is used every element is printed twice, so \texttt{[k k r r e e g g]} is four elements and eight printed symbols.

\footnotesize\textsuperscript{b}A nine-element step-two sequence spans $17$ alphabet positions, so expert items use the $26$-letter alphabet rather than the $15$-symbol set. At the shorter levels either alphabet fits, so the alphabet varies alongside the axes rather than independently of them, and it is not one of the counted axes.

\footnotesize\textsuperscript{c}The alphabet is drawn per item. Two of the three combination items use the $15$-symbol set and one the $26$-letter set. Their third counted generalization axis is the two-rule composition itself rather than one of the axes above.
\end{table}

\paragraph{The ambiguity verifier}
A single worked example under-determines the rule, e.g.\ when considering a sort item where the
example is equally consistent with ``sort the whole sequence into alphabet
order'' and with ``swap the elements at positions $i$ and $j$''. These agree on
the example and disagree on the target. Showing a second example only helps if
it differs on the disambiguating feature, and two randomly drawn examples can
share the same accident and leave the ambiguity intact.

The generator therefore constructs a hypothesis set $\mathcal{H}$ of rules a
solver might plausibly induce, including the intended structural families parameterised
over step sizes $1$--$4$, plus enumerated positional swaps, positional removals,
per-position ``fix the odd one out'' readings, alphabet-order sorting, duplicate
removal, arithmetic-progression repair, and a literal per-position index-shift
reading of each shown example. An item is accepted only if \emph{every} rule in
$\mathcal{H}$ that is consistent with \emph{all} shown examples maps the target
to the same intended answer, and is resampled otherwise. After generation, a
self-check re-parses every emitted sequence-based item from its stored display strings and
re-runs the verification independently of the in-memory state used during
generation, asserting that the guarantee holds. 
Verification is relative to the enumerated hypothesis families in $\mathcal{H}$. Every positional rule in $\mathcal{H}$ is indexed by position, leaving content-addressed rules such as ``swap whichever adjacent pair is out of order'' outside the hypothesis set. The two exceptions below illustrate this limit.

\paragraph{The combination level}
Three items apply two transformations at once (remove-redundant $+$ sort, remove-redundant $+$ add-letter, sort $+$ add-letter). For these the hypothesis set extends to all ordered two-operation compositions of a curated primitive set, so the verifier can represent the intended pair and rule out confusable ones. The extension is gated behind a flag so single-operation levels keep their original hypothesis set and their generated output byte-identical.

\paragraph{Accepted alternate answers}
Two combination-level items were genuinely under-determined in a way the verifier did not catch. Each worked example needed only one adjacent-pair swap to reach the intended order, so ``sort into alphabet order'' and ``swap the single out-of-order pair'' fit both examples equally well, and only the target distinguishes them. For these two items the answer implied by the narrower rule is also accepted as correct.

\section{Questionnaire sections and attention checks}
\label{app:questionnaire}

Four post-task questionnaire sections were administered. \emph{About you} (all participants) covered age, gender, education, profession, frequency of AI-tool use ($5$-point, \emph{Never}--\emph{Every day}), and preferred AI provider and model. \emph{AI consultation strategy} (\textit{AI} only) asked for the percentage of items with interaction beyond the mandated minimum, a four-option description of the typical approach, and behavior when the assistant disagreed with the participant's initial answer --- a self-report companion to the behavioral reliance measure. \emph{AI interaction} (\textit{AI} only) rated perceived helpfulness, trust, and frustration on fully labeled $5$-point scales. \emph{Task feedback} (all) rated the task set's suitability and collected optional strategy descriptions and free-text comments. Full item wording is included with the study materials (see the availability note opening Section~\ref{sec:results}).

Three attention checks required the instructions to have been read rather than skimmed. The first asked how the $\pounds100$ bonus is earned, with the strongest distractor being the \emph{other} bonus in the same incentive card. A second condition-specific check probed the engagement disclaimer (\textit{AI}) or the unguessable size of the top-10 bonus (\textit{no-AI}). The third was a standard instructed-response item after the battery (``select the leftmost option''), carried by a condition-appropriate scale. Exclusions applied to participants failing two or more (Section~\ref{sec:method:participants}). In the analyzed sample no one failed more than one, by construction. Of the 535 participants, 459 passed all three, and the 76 single failures fell mostly on the first check (60), whose strongest distractor is the other bonus in the same incentive card.

\section{Model benchmarking harness}
\label{app:harness}

Section~\ref{sec:method:benchmark} gives the scoring and exclusion rules. Eighteen claude-opus-4-8 letter-string replies were stopped by a provider-side safety classifier. 32 replies carried needs-review flags, overlapping on 13 replies, so 37 of 16{,}000 had at least one flag. Custom-alphabet prompts appeared to trigger a cipher/jailbreak classifier, truncating or suppressing replies. This occurred systematically with claude-opus-5 in early testing but emerged for claude-opus-4-8 only in the 100-run campaign, not its ten-run screening.

All $32$ needs-review rows and all $18$ refusals received incorrect scores before exclusions. The needs-review rows were then excluded, including the $13$ that also carried refusal flags.
The influence is bounded. The ordering of the four assistants is identical whether the needs-review rows are kept, dropped, or all flipped to correct, and no assistant's competence moves by more than $.004$ across those three treatments. Excluding the refusals instead of scoring them would move the affected group's pooled competence by $.002$.

The identified flags fall $15$ / $12$ / $5$ / $0$ across claude-opus-4-8, gpt-5.6-luna, kimi-k3 and gemini-3.6-flash. The sensitivity checks above assess these flagged replies.

\paragraph{Message parity.}
\label{sec:method:benchmark:parity}
A model that is briefed better than a human is not a fair reference. The chat-parity harness that produced the reported numbers therefore sends the assistant exactly the messages a participant's browser sent and no system prompt, no output-format constraint and no prefill. Benchmark history accumulates within a task and resets between tasks. The runs replay each item's stimulus message rather than participants' follow-up turns, whose effect Section~\ref{sec:groundtruth} bounds (advice accuracy .713 to .751 after revision). The benchmark campaign ran 13--19 August 2026, after all participant sessions (22 July--12 August).
Two residual asymmetries favor the humans. Participants receive an instruction page and a practice trial per task that the assistant does not, and the assistant's conversation history is reset at each task, whereas participants' chat history accumulated across the whole session.

We used gemini-3.5-flash-lite with thinking disabled for both the benchmark answer-mapping described above and the transcript extraction in \autoref{sec:method:reliance}. Note that this makes the two pipelines share a failure mode.

\paragraph{Run bookkeeping and robustness.}
Each run is written to its own timestamped directory with an explicit run identifier. Error-flagged runs are dropped from the average. In the chat-parity campaign two runs aborted mid-way (one from a provider spending cap and one from an empty reply that poisoned the conversation history) and were re-run cleanly, with only the discarded partial attempts quarantined.

\section{Sample composition}
\label{app:armcomp}

\autoref{tab:armcomp} reports the analyzed sample's composition per group.

\begin{table}[tb]
\centering
\caption{Composition of the analyzed sample by group. Age is $M$ ($SD$). Remaining columns are percentages. Other combines non-binary, multiple categories, and undisclosed gender. AI use is the share reporting at least weekly, and daily, use. Provider is the share naming OpenAI as preferred provider.}
\label{tab:armcomp}
\small
\begin{tabular}{lrrrrrrrrr}
\toprule
Group & $n$ & Age & Men & Women & Other & Bachelor+ & Weekly+ & Daily & OpenAI pref. \\
\midrule
unaided            & 187 & 36.8 (10.9) & 53 & 47 & 1 & 73 & 77 & 37 & 55 \\
gpt-5.6-luna       & 179 & 34.3 (10.8) & 54 & 45 & 1 & 68 & 84 & 47 & 56 \\
claude-opus-4-8    &  61 & 37.7 (10.8) & 52 & 48 & 0 & 77 & 85 & 54 & 56 \\
gemini-3.6-flash   &  52 & 33.0 (10.0) & 33 & 67 & 0 & 71 & 79 & 42 & 67 \\
kimi-k3            &  56 & 32.3 (10.4) & 45 & 52 & 4 & 71 & 79 & 43 & 50 \\
\bottomrule
\end{tabular}
\end{table}

\section{Supplementary results}
\label{app:supp}

\subsection{Robustness checks}

\paragraph{Input and sampling uncertainty.} Resampling participants within groups and paired benchmark runs 4{,}000 times gave fixed-battery $S=-.062$ [$-.080$, $-.046$] and pooled capture $.587$ [$.527$, $.645$]. These percentile intervals propagate component uncertainty on the same 40 items; they differ from the hierarchical-model intervals. Stricter-sampler refits of $S$, capture, the item-level competence gradient, and the ladder retained data, formulas, and priors. Each had 1--5 divergences in 40{,}000 draws. Point estimates were stable, but capture's 95\% HDI widened from $[.356,.809]$ to $[.317,.839]$. Full comparisons accompany the supplement.

\paragraph{Anisotropy.} The two sides of that boundary are not mirror images. Each unit by which the assistant leads the unaided participant is worth $.578$ [$.473$, $.686$] of team accuracy, while each unit by which it trails costs $-.293$ [$-.445$, $-.151$], with a benefit-minus-harm magnitude contrast of $.285$ [$.069$, $.501$]. Harm is real and resolves, but accrues at about half the rate the benefit does. The team is more responsive to what the assistant knows than to what it does not.

\paragraph{Pass-through on the realized-advice axis.} Benchmark competence is a noisy proxy for delivered advice, which can also change through participant follow-ups. Refitting the per-assistant slopes on the realized-advice axis --- the share of that group's trials with correct advice on the item, same trial-level specification on both axes --- gives $4.308$ [$3.870$, $4.765$] for gpt-5.6-luna, $3.829$ [$3.325$, $4.324$] for claude-opus-4-8, $3.747$ [$3.216$, $4.300$] for gemini-3.6-flash and $3.156$ [$2.447$, $3.857$] for kimi-k3. The spread narrows to roughly $1.4\times$, and only the gpt-5.6-luna--kimi-k3 contrast resolves ($1.150$ [$0.471$, $1.835$]), where on benchmark competence every kimi-k3 contrast does. Changing axes changes both the predictor's operationalization and its measurement error.

Benchmark competence is a noisy stand-in for the advice participants actually met. Fitting the same trial-level model on each of two competence axes in turn --- the benchmark, then what the assistant actually said in this study's own transcripts, which correlate at $.898$ --- steepens the slope from $2.534$ [$2.245$, $2.829$] to $4.208$ [$3.848$, $4.584$]. Both are separate fits from the pooled pass-through model above, so the three slopes are not competing estimates of one quantity.

\paragraph{Task-order adjustment.} Adjusting for task order moves the trial-level competence slope from $2.133$ [$1.046$, $3.042$] to $2.124$ [$1.010$, $3.004$] log-odds per unit. This adjusts for position in the session, not task identity. Separately, competence slopes were positive within all four tasks (Section~\ref{sec:rq1}).

\paragraph{Censoring robustness.} Competence is right-censored, with 38 of the 160 cells at exactly 1.00, but the outcome ceiling does not bind. No item leaves the unaided group above .90, and the room to improve is flat across competence, so the rising effect is not an artefact of shrinking headroom.

\subsection{The full policy ladder}

This section reports the levels behind the contrasts in \autoref{fig:main}B. The model has 1{,}120 rows, comprising 160 item--assistant cells crossed with seven reference levels (unaided, observed assisted, first answer, benchmark, task-aware, item-aware, and independent-error oracle). A Gaussian model estimates level-specific means with random intercepts for item and assistant. The task/item references choose between the separate unaided group's accuracy and benchmark competence.

\par First-answer accuracy uses the earliest extracted assistant answer and covers about 97\% of assisted trials, whereas observed performance includes all assisted trials. These levels have different eligibility and are reported descriptively.

\par The fitted unaided level is .525 [.432, .613], observed assisted accuracy .726 [.634, .814], and the conditional first-answer reference .727 [.634, .814]. The task-aware reference is .761 [.670, .851], the item-aware reference .787 [.696, .876], and the independent-error oracle .867 [.776, .956]. The item-aware and oracle gains over observed performance are $.061$ [$.030$, $.092$] and $.141$ [$.110$, $.172$]. These model-based reference contrasts use the component-accuracy estimates and assumptions specified above.

\par Of the 348 assisted participants, the 88 who estimated that the assistant was no better than themselves reached .745 [.733, .758], versus .713 [.705, .721] among the other 260, a difference of .032 [.017, .046]. These post-task beliefs may reflect experienced performance. Mean estimated assistant performance was 29.7 against 28.8 correct answers out of 40. This comparison concerns mean beliefs and mean performance.

\subsection{Metacognition across the four assistants}

This section reports per-assistant item-level detail for \autoref{sec:rq3}. The per-group confidence gaps and their drops against unaided are in \autoref{tab:confgap}.
\par Across the 40 items, correlations between mean confidence and LLM competence were $r = -.363$ for gpt-5.6-luna, $-.383$ for claude-opus-4-8, $-.502$ for gemini-3.6-flash, and $-.613$ for kimi-k3. Competence correlated $.105$ with unaided accuracy, and adjusting for that measure strengthened the negative associations. The pattern arose between tasks and was inconsistent within them, where two tasks offered little competence variation. Mean post-advice confidence therefore did not consistently track item-level LLM competence. This differs from discriminating the correctness of one's submitted answer and leaves pre-advice confidence untested.

\end{document}